\documentclass[12pt]{article}
\pdfoutput=1
\usepackage[]{hyperref}
\usepackage[]{putex}
\usepackage[utf8]{inputenc}
\usepackage[table]{xcolor}
\usepackage{amsmath,amsthm,amssymb}
\usepackage{tikz}
\usepackage{tikz-cd}
\usetikzlibrary{decorations.markings}
\usetikzlibrary{calc}
\usepackage{subcaption}
\usepackage{enumerate}
\usepackage{enumitem}
\usepackage{multirow}
\usepackage{mathtools}
\usepackage{bm}
\usepackage{cite}
\allowdisplaybreaks

\newcommand{\beq}{\begin{equation}}
\newcommand{\eeq}{\end{equation}}
\newcommand{\bea}{\begin{eqnarray}}
\newcommand{\eea}{\end{eqnarray}}

\definecolor{colour1}{rgb}{0.8, 0.6, 1}
\definecolor{cyan2}{rgb}{0, 1, 1}
\definecolor{green2}{rgb}{0.2, 0.5, 0.2}

\definecolor{colour1}{rgb}{0.3, 0.3, 1}
\definecolor{colour2}{rgb}{1, 0, 0}
\definecolor{colour3}{rgb}{0.5, 0.9, 1}
\definecolor{colour4}{rgb}{0.7, 0.4, 0.8}
\definecolor{colour5}{rgb}{1, 0.5, 0.2}
\definecolor{colour6}{rgb}{0, 0.7, 0}
\definecolor{colour7}{rgb}{0.75, 0.9, 0.75}
\definecolor{colour8}{rgb}{0.8, 0.6, 0.6}
\definecolor{colour9}{rgb}{1, 0.6, 1}

\usepackage{setspace}
\allowdisplaybreaks

\begin{document}

\vskip-2cm
\title{Finite Temperature at Finite Places}

\authors {An Huang$^1$ \& Christian Baadsgaard Jepsen$^2$}
\institution{Brandeis}{
$^1$Department of Mathematics, Brandeis University, Waltham, MA 02453, USA.
}
\institution{KIAS}{$^2$School of Physics, Korea Institute for Advanced Study, Seoul 02455, Korea.}

\date{\today}

\abstract{
This paper studies AdS/CFT in its $p$-adic version (at the ``finite place") in the setting where the bulk geometry is made up of the Tate curve, a discrete quotient of the Bruhat-Tits tree. Generalizing a classic result due to Zabrodin, the boundary dual of the free massive bulk theory is explicitly derived. Introducing perturbative interactions, the Witten diagrams for the two-point and three-point correlators are computed for generic scaling dimensions at one-loop and tree level respectively. The answers obtained demonstrate how $p$-adic AdS/CFT on the Tate curve provides a useful toy model for real CFTs at finite temperature.
}

\maketitle
{\setstretch{1.25}
\tableofcontents
}
\section{Introduction and Summary}
\label{sec:Introduction}

In modelling a physical system, theoretical predictions sometimes remain robust even as we change the details of the model. This makes life as a theorist easier because it justifies us in neglecting extraneous complications and picking the simplest conceivable theory when endeavouring to understand the system. Even when precise quantitative data is lost in the choice of a simple model, it can still be possible to extract valuable qualitative information. The kinds of simplifications made for the sake of computational tractability can sometimes be rather drastic, like studying $SU(N)$ Yang-Mills theory with $N$ taken to infinity in place of $SU(3)$, or computing path integrals for quantum gravity in two dimensions instead of four. Perhaps the most extreme instances of this practice are those cases where the underlying number field of a theory has been chosen not to be the real numbers $\mathbb{R}$ or the complex numbers $\mathbb{C}$ but rather some other field like the $p$-adic numbers $\mathbb{Q}_p$. The possibility, under the right circumstances, to subject a theory to such a fundamental alteration and yet still preserve its key features has been attested by a number of studies of $p$-adic string theory \cite{Volovich:1987nq,Grossman:1987up,Freund:1987kt} and $p$-adic AdS/CFT \cite{Gubser:2016guj,Heydeman:2016ldy}, although it is often necessary to choose one particular version among several equivalent formulations of equations in real theories in order to observe a manifest parallelism to analogous equations in $p$-adic theories. Examples of this phenomenon include the following:

--- The familiar CFT two- and three-point functions assume identical forms in the $p$-adic setting, except with vector norms replaced with $p$-adic norms \cite{Gubser:2016guj}, and the decomposition of four-point correlators into conformal blocks too admits a bulk dual description in terms of geodesic Witten diagrams \cite{Gubser:2017tsi}. If one goes to Mellin space, the  Witten diagrams of real and $p$-adic AdS/CFT admit near exact parallel descriptions at any number of points provided the Mellin amplitudes are expressed via a suitable contour representation \cite{Jepsen:2018dqp,Jepsen:2018ajn}. 

--- Perturbative computations of the anomalous dimensions for the $p$-adic $O(N)$ model at small $\epsilon$ as well as at large $N$ closely mirror the real calculations \cite{Gubser:2017vgc}.

--- It is possible to formulate a $p$-adic version of entropy that obeys strong subadditivity and monogamy of mutual information as well as a version of the Ryu-Takyanagi formula \cite{Heydeman:2018qty}.

--- The $p$-adic analog of the torus partition function exhibits a structure reminiscent of the partition function of a complex CFT with diagonal modular invariant \cite{Hung:2019zsk}.

The vast majority of work on $p$-adic theories has sought to replicate corresponding results in real theories and to uncover the similarities and differences that exist across the different types of theories. But in a small number of instances, the $p$-adic computations were carried out first and helped to illuminate features of real theories:
\begin{itemize}
    \item  Ref.~\cite{Brekke:1987ptq} succeeded in evaluating the $p$-adic Koba-Nielsen amplitude for any number of tachyons and to determine the associated effective spacetime action, which for real string theory was determined later on in Ref.~\cite{Gerasimov:2000zp}. The spacetime action from $p$-adic string theory provided an instructive toy model for the study of tachyon condensation in string field theory \cite{Ghoshal:2000dd,Minahan:2001wh,Sen:2004nf}.
    \item The holographic duals of the five- and six-point conformal blocks, along with the conformal block decompositions of Witten diagrams into these blocks and a set of propagator identities instrumental in performing these decompositions were all computed on the $p$-adic side first, before the calculations were uplifted to the real place \cite{Gubser:2017tsi,Parikh:2019ygo,Jepsen:2019svc}. 
\end{itemize}

A key computational advantage to studying $p$-adic theories is that the derivative operator is entirely absent, for the reason that, in the customary formulations of $p$-adic physics, fields are still taken to be real- or complex-valued even though their arguments are $p$-adic. The absence of derivative operators implies the absence, not just of all descendants, but also of all higher-twist families of operators, which entails tremendous simplifications as infinite sums and complicated hypergeometric functions collapse down to a small number of terms.

In the present paper, we undertake to widen the scope of $p$-adic AdS/CFT by bringing it to also encompass CFTs at finite temperature.\footnote{Since adjectives are not associative, perhaps it behoves us to clarify that the subject of this paper is $p$-adic (CFTs at finite temperature), ie. the $p$-adic AdS/CFT analog of finite temperature CFTs. The subject of ($p$-adic CFTs) at finite temperature, as applicable to hierarchical spin chains at finite temperature, similar to those measured in \cite{Periwal:2021eur}, is quite a different animal, since in this case the compact thermal direction remains real so that the field theory description contains fields with mixed real and $p$-adic arguments as in Ref.~\cite{Gubser:2018ath}.} \footnote{ Other recent works related to $p$-adic AdS/CFT include \cite{Hung:2018mcn,Qu:2019tyi,Garcia-Compean:2019jvk,Huang:2021bsg,Qu:2021huo,Qu:2021fgz,Costantino:2020vdu,Huang:2020aao,Huang:2020vwx,Garcia-Compean:2022pjw,Delporte:2023saj,Okunishi:2023syy,Qu:2024hqq}.}

\subsection{Thermal CFTs at the real place}
At the real place, thermal CFTs are CFTs that live on the manifold\footnote{Since $S^1 \times \mathbb{R}$ is conformal to the plane, thermal CFT correlators in two real dimensions can be obtained from zero-temperature correlators by a conformal transformation. At the real place, the interesting cases to consider are therefore those with $d>2$.} $S^1\times \mathbb{R}^{d-1}$. The present understanding of such CFTs grew out of early important work on finite-temperature two-point functions, including Refs.~\cite{Sachdev:1992py,Chubukov:1993aau,Balasubramanian:1999zv,Louko:2000tp}, and the essential properties of the correlators as determined by the underlying symmetry of these theories were laid down in the foundational work of Refs.~\cite{El-Showk:2011yvt,Katz:2014rla,Witczak-Krempa:2015pia,iliesiu2018conformal}. But despite experimental applications within the realm of quantum phase transitions, the literature on thermal CFTs lags far behind the knowledge of zero-temperature CFTs, with existing results on conformal blocks, conformal block decompositions, and Witten diagrams mainly restricted to one- and two-point correlators, although the field is continually witnessing exciting progress, as evidenced by Refs.~\cite{Iliesiu:2018zlz,Gobeil:2018fzy,Manenti:2019wxs,Ghosh:2019sqf,Dodelson:2020lal,Alday:2020eua,Grinberg:2020fdj,Rodriguez-Gomez:2021pfh,Krishna:2021fus,Rodriguez-Gomez:2021mkk,Petkou:2021zhg,Karlsson:2022osn,Dodelson:2022yvn,David:2022nfn,Georgiou:2022ekc,Bhatta:2022wga,Diatlyk:2023msc,Dodelson:2023vrw,Marchetto:2023xap,Bhatta:2023qcl,He:2023wcs,Georgiou:2023xpg,Esper:2023jeq,Marchetto:2023fcw,Bobev:2023ggk,Parisini:2023nbd,Allameh:2024qqp,Alkalaev:2024jxh,Ceplak:2024bja,Barrat:2024}. There are no fundamental hindrances to extending the known thermal CFT results to wider classes of observables, in particular higher-point correlators, but progress is hindered by the shear technical difficulty of carrying out computations. Repeatedly applying the operator product expansion (OPE) to any correlator for whose insertions the OPE converges, the correlator ultimately reduces to a sum of one-point functions. At zero temperature, this sum collapses on the identity operator, which alone has a non-zero CFT one-point function, but in the presence of a finite temperature, which introduces an energy scale to the system, the final sum remains unwieldy, as any traceless symmetric operator can have a non-zero one-point function. To help alleviate this obstruction, $p$-adic AdS/CFT provides a simplified playground in which to study the skeleton for the body of calculations that have proven to be particularly challenging to perform over the reals.

\begin{figure}
    \centering
\begin{align*}
\begin{matrix}
\\[-41pt]
\text{
\includegraphics[scale=0.8]{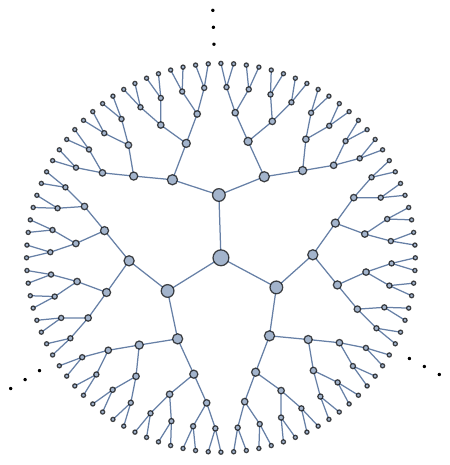}}
\end{matrix}
\hspace{25mm}
\begin{matrix}\\[-9mm]
\text{
\includegraphics[scale=0.63]{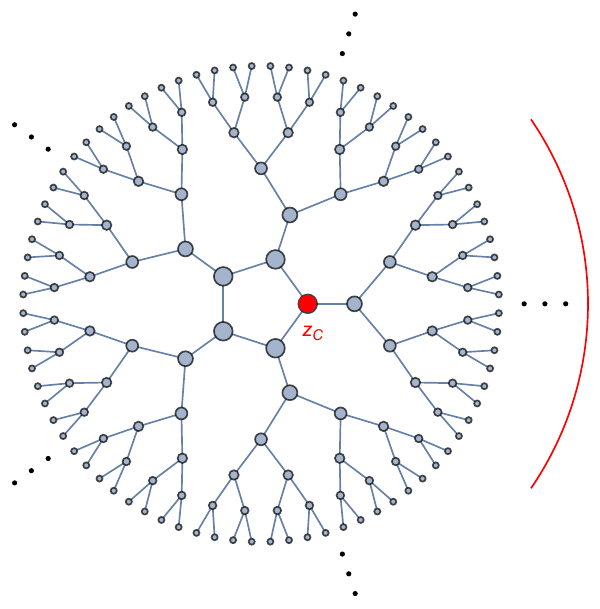}}
\end{matrix}
\end{align*}
    \caption{Left: The Bruhat-Tits tree $T_{p^d}$ with $p^d=2$. Right: The Tate curve $T_{p^d}^{(w)}$ with $p^d=2$ and $w=5$. To any vertex $z_C$ on the thermal cycle on the Tate curve, we associate a subset of the boundary points as indicated by the red curve.}
    \label{fig:geometries}
\end{figure}

\subsection{Thermal CFTs at the finite places}
\label{subSec:thermalFinite}
The bulk duals to real finite-temperature CFTs are furnished by black holes or black branes in AdS, and in the limit of vanishing gravitational interactions the bulk space is given by thermal AdS. This space can be constructed by performing a quotient of anti-de Sitter space with respect to a discrete subgroup of the isometry group. In the $p$-adic setting, the zero-temperature bulk space, often called the Bruhat-Tits tree, is the infinite regular tree of degree $p^d+1$, where $p$ is a prime number and $d$ is a natural number, which enters various formulas in a similar manner to way the number of spacetime dimensions appears in formulas in real AdS/CFT. The analog to the finite-temperature bulk space at the $p$-adic place is similarly constructed by a discrete quotient \cite{Manin:2002hn,Heydeman:2016ldy}. For any $w\in \mathbb{N}$ the quotient is constructed by picking any bi-infinite path through the tree and identifying all vertices on this path that are separated by $w$ edges. The resulting bulk geometry is sometimes referred to as the Tate curve $T_{p^d}^{(w)}$. 
Figure \ref{fig:geometries} shows an example of a Bruhat-Tits tree along with a Tate curve constructed from it. The graph that constitutes the Tate curve contains a single cycle. We will refer to this cycle, which contains $w$ edges, as the thermal cycle. 

Several papers \cite{Heydeman:2016ldy,Heydeman:2018qty,Marcolli:2018ohd,Ebert:2019src,Chen:2021rsy,Chen:2021qah,Yan:2023lmj} in the $p$-adic AdS/CFT literature have studied theories on the Tate curve\footnote{See also \cite{Hung:2019zsk} for a study of partition functions on this geometry from a tensor network point of view.} from a bulk perspective and further argued that this geometry furnishes the $p$-adic analog of the Euclidean BTZ black hole, with the parameter $w$ providing the analog of black hole entropy and area.\footnote{In light of the related themes of Ref.~\cite{Ebert:2019src} and the present paper, it may be useful to note that the former sought to extend the results of Ref.~\cite{Kraus:2016nwo} to $p$-adic AdS/CFT, while the present paper derives inspiration mainly from Refs.~\cite{iliesiu2018conformal} and \cite{Alday:2020eua}.} 
While Schwarzschild-AdS and thermal AdS furnish distinct bulk geometries at the real place, we argue in this paper that the selfsame Bruhat-Tits quotient that in the literature has been interpreted as a kind of BTZ black hole also provides a bulk analog of thermal AdS, as we will demonstrate by exhibiting a parallelism in the computations of holographic boundary correlators at the real and finite places.

But in order to display this parallelism, we will need to perform an identification of what at the finite place constitutes the unique dimensionful scale present in the system. In real finite-temperature CFTs, the inverse temperature is the only scale in the system. In consequence, one can set it equal to one and always be able to restore it later on via dimensional analysis. The analogous parameter for the Tate curve is set by the volume of the boundary points. To be more precise, let $z_C$ be one of the $w$ vertices on the thermal cycle, let $x$ be a boundary point, and let $z_C\rightarrow x$ signify that there is a semi-infinite path from $z_C$ to $x$ that has no edges on the thermal cycle. The set of such points $x$ are indicated by the red arc in Figure~\ref{fig:geometries}. The Tate curve analog to inverse temperature is the scale set by the normalization of the boundary integration measure:
\begin{align}
\label{measureNormalization}
\int_{z_C\,\rightarrow\,x}dx = \frac{p^d-1}{p^d}\,|\beta|^d\,.
\end{align}

The set of boundary points can be identified with a discrete quotient of the degree $d$ unramified extension $\mathbb{Q}_{p^d}$ of the $p$-adic numbers,
\begin{align}
\label{pnary}
x = p^{v(x)} \sum_{n=0}^\infty\,a_n\,p^n\,,
\hspace{10mm}
\text{where}
\hspace{10mm}
a_n\in \mathbb{F}_{p^d}\,,
\hspace{5mm} a_0\neq 0\,,
\end{align}
where performing the quotient means that the integer $v(x)$, known as the valuation of $x$, is identified mod $w$ and so is valued over $\mathbb{Z}_w$ instead of $\mathbb{Z}$. In other words, $\partial T_{p^d}^{(w)}= \mathbb{Q}_{p^d}/(w\mathbb{Z})$.

The integration measure in \eqref{measureNormalization} is equal to $|\beta|^d$ times the multiplicative Haar measure on $\mathbb{Q}_{p^d}$. An equivalent interpretation of the set of boundary points is to think of any boundary point $x$ as being uniquely specified by a doublet $(v,\mathsf{x})$ where $v\in \mathbb{Z}_w$ and $\mathsf{x}\in \mathbb \beta \mathbb{U}_{p^d}$, where $\mathbb{U}_{p^d}$ is the subset of points in $\mathbb{Q}_{p^d}$ that have unit norm. The $p$-adic norm $|\cdot|_p$ of a number $x\in\mathbb{Q}_{p^d}$ given as in \eqref{pnary} is defined as $|x|_p=p^{-v}$, but as this norm does not respect the equivalence $v\sim v+w$, we will be using a slightly different norm $|\cdot |$. This norm $|\cdot|$ will be used to measure the distance only between pairs of boundary points that have the same valuation $v$. Given two boundary points $x=(v,\mathsf{x})$ and $y=(v,\mathsf{y})$, we define the norm of their difference as
\begin{align}
|x-y|= |\mathsf{x}-\mathsf{y}|_p= |\beta|\,\frac{|x-y|_p}{|x|_p}= |\beta|\,\frac{|x-y|_p}{|y|_p}\,.
\end{align}
With the definitions of $|\beta|$ and the boundary norm in place, it will be possible for us to uncover a number of formulas in $p$-adic holography that parallel formulas at the real place. 

We have identified $|\beta|$ as the unique scale of the $p$-adic finite temperature CFT but we refrain from referring to it as inverse temperature. The isometry group of the Tate curve $T_{p^d}^{(w)}$ includes a factor of $\mathbb{Z}_w$ that rotates the branches on the thermal cycle into each other, and, in analogy with finite temperature at the real place, it is natural to expect any notion of $p$-adic temperature to be related to this cyclic group and to depend on $w$. One could include factors of $w$ or $p^w$, say, in the definition of $|\beta|$ so as to introduce a dependency on the cyclic group, but the scope of our work provides no way of singling out one such definition as most natural.

\subsection{Parallel formulas}
Let us consider the two-point boundary correlator of two identical scalars in the two formalisms. For two boundary points $x_1=(\tau_1,{\bf x}_1)$ and $x_2=(\tau_2,{\bf x}_2)$ in real thermal holography on $S_\beta^1\times \mathbb{R}^{d-1}$, the operator product expansion converges when the condition is satisfied that
\begin{align}
\label{convergenceCondition}
(\tau_1-\tau_2)^2+({\bf x}_1-{\bf x}_2)^2< \beta^2\,.
\end{align}
If so, the two-point function of two identical scalars with scaling dimension $\Delta$ in mean field theory admits the conformal block expansion determined in Ref.~\cite{iliesiu2018conformal} to be given as follows, where we use a superscript $(0)$ to indicate the absence of interactions,
\begin{align}
\label{realOO}
&\left<\mathcal{O}_\Delta(x_1)\mathcal{O}_\Delta(x_2)\right>^{(0)}=\frac{1}{|x_1-x_2|^{2\Delta}}
+\sum_{n\in\mathbb{N}_0} \sum_{\ell \in 2\mathbb{N}_0}
a_{[\phi\phi]_{n,\ell}}
\,C_\ell^{(\nu)}(\eta)\,
\frac{|x_1-x_2|^{2n+\ell}}{|\beta|^{2\Delta+2n+\ell}}\,, 
\\[7pt] \nonumber
&\text{where}
\hspace{6mm}a_{[\phi\phi]_{n,\ell}}=2\zeta(2\Delta+2n+\ell)
\frac{(\ell+\nu)(\Delta)_{\ell+n}(\Delta-\nu)_n}{n!(\nu)_{\ell+n+1}}\,,
\hspace{6mm}
\nu=\frac{d-2}{d}\,,
\hspace{6mm}
\eta=\frac{\tau_1-\tau_2}{|x_1-x_2|}\,.
\end{align}
In the above formulas, $C^{(\nu)}_\ell(\eta)$ are Gegenbauer polynomials, and $(\Delta)_n = \Gamma(\Delta+n)/\Gamma(\Delta)$ is the Pochhammer symbol. Since derivative contributions will be absent in the analogous $p$-adic formula, it is instructive to see the real formula for the correlator when only contributions from derivative-free operators are explicitly included:
\begin{align}
\label{OOreal}
\left<\mathcal{O}_\Delta(\tau_1,x_1)\mathcal{O}_\Delta(\tau_2,x_2)\right>^{(0)}\hspace{-0.5mm}=\hspace{-0.5mm}
\frac{1}{|x_1-x_2|^{2\Delta}}
\hspace{-0.5mm}+\hspace{-0.5mm}
\frac{2\zeta(\Delta)}{|\beta|^{2\Delta}}+\text{(terms from ops. with derivatives)}\,.
\end{align}

In $p$-adic AdS/CFT, the analog of the convergence condition \ref{convergenceCondition} for two boundary points $x_1$ and $x_2$ is the condition that $x_1$ and $x_2$ emanate from the same vertex on the thermal circle. In other words, there exists a vertex $z_C$ on the thermal circle such that $z_C \rightarrow x_1$ and $z_C \rightarrow x_2$. From the boundary point of view, the condition amounts to the stipulation that $x_1$ and $x_2$ have the same valuation, $v(x_1)=v(x_2)$. When this condition is fulfilled, the analogous $p$-adic two point correlator, first computed in \cite{Heydeman:2016ldy} using a different notation, turns out to be given as
\begin{align}
\label{OO}
\left<\mathcal{O}_\Delta(x_1)\mathcal{O}_\Delta(x_2)\right>^{(0)}
=\frac{1}{|x_1-x_2|^{2\Delta}}
+\frac{2\zeta_p(w\Delta)p^{-w\Delta}}{|\beta|^{2\Delta}}\,,
\hspace{8mm}
\text{where}
\hspace{8mm}
\zeta_p(s) = \frac{1}{1-p^{-s}}\,.
\end{align}
The two correlators \eqref{realOO} and \eqref{OO} are obtained holographically from tree-level Witten diagrams. We can also compare the one-loop diagrams. In a bulk theory with a perturbative quartic interaction, the one-loop two-point correlator was determined in Ref.~\cite{Alday:2020eua} to be given by
\begin{align}
\label{OO1real}
\left<\mathcal{O}_\Delta(x_1)\mathcal{O}_\Delta(x_2)\right>^{(1)}
=\,&
-\frac{2^{-d-2}\pi^{-d/2}}{\Gamma(\Delta)\Gamma(\Delta-\frac{d-2}{2})}
\int_{s_0-i\infty}^{s_0+i\infty}
\frac{ds}{2\pi i}
\Gamma(s)^2\Gamma(\Delta-s)^2 \times
\\[5pt] \nonumber
&
\frac{\zeta(2\Delta-2s)\Gamma(\Delta-\frac{d-1}{2}-s)\Gamma(2\Delta-\frac{d}{2}-s)}{\Gamma(\Delta+\frac{1}{2}-s)\Gamma(2\Delta-d+1-s)}
\frac{\left<\mathcal{O}_s(x_1)\mathcal{O}_s(x_2)\right>^{(0)}}{|\beta|^{2\Delta-2s}}\,,
\end{align}
where $s_0$ is any complex number with $0 < \text{Re}[s_0] < \Delta-\frac{d-1}{2}$. At the $p$-adic place, the analogous one-loop computation produces an answer that can be expressed via a similar contour integral as
\begin{align}
\label{OO1}
\left<\mathcal{O}_\Delta(x_1)\mathcal{O}_\Delta(x_2)\right>^{(1)}
=\,& -p^{-w\Delta}\zeta_p(w\Delta)\,
\frac{\zeta_p(4\Delta-d)}{\zeta_p(2\Delta)^2} \times
\\ \nonumber
&
\frac{\log p}{2\pi i}
\int_{s_0-\frac{i\pi}{\log p}}^{s_0+\frac{i\pi}{\log p}}ds\,\,
\frac{\zeta_p(\Delta-s)^2\zeta_p(2s)^2}{\zeta_p(2\Delta+2s)}\,
\frac{\left<\mathcal{O}_s(x_1)\mathcal{O}_s(x_2)\right>^{(0)}}{|\beta|^{2\Delta-2s}}\,,
\end{align}
where $s_0$ is any complex number with $0 < \text{Re}[s_0] < \Delta$.

As $p$-adic correlators are computed with more facility than their real versions, we are able too to carry out a number of computations, for which the real counterparts have not yet been performed, and for which the answers may help to guide or motivate future real progress. Specifically, in this paper, we present also the generalization of \eqref{OO1} for non-identical scaling dimensions as well as the $p$-adic conformal block decomposition for the mean field three-point correlator.

\subsection{Outline of paper}
The remainder of this paper is organized as follows:

Section~\ref{sec:Dirichlet} describes the free bulk lattice action on the Tate curve and the solutions to the equation of motion and presents the requisite types of boundary conditions to impose on the system, along with the general solution to the bulk equation of motion subject to any such boundary conditions.

Section~\ref{sec:Boundary} utilizes the results of the previous section to evaluate the free partition function as a functional of the boundary conditions, thereby obtaining the holographic boundary dual action to the free massive bulk action.

Section~\ref{sec:TwoPoint} reviews the mean field two-point function for a pair of identical scalar operators and also presents the one-loop two-point function in the presence of a perturbative quartic bulk interaction between scalars of arbitrary identical or distinct scaling dimensions.

Section~\ref{sec:ThreePoint} presents the tree-level holographic three-point function.

Section~\ref{sec:Outlook} concludes the main portion of the paper with speculations on possible future work.

Appendix~\ref{Appendix} provides the detailed calculations behind most of the results of the paper.

\section{The Dirichlet Problem on the Tate Curve}
\label{sec:Dirichlet}

Given a function $\phi(z)$ that takes arguments on the vertices $z$ of a graph, the graph Laplacian $\Box$ is defined as
\begin{align}
\Box \phi(z) = \sum_{z'\sim z}\Big(\phi(z')-\phi(z)\Big)\,,
\end{align}
where the sum runs over all vertices $z'$ that are adjacent to $z$. This operator can be used to write down a free massive action on generic graphs: 
\begin{align}
S = -\frac{\kappa}{2}\sum_{z}\phi(z)(\Box-m^2)\phi(z)\,.
\end{align}
The associated equation of motion is given by
\begin{align}
(\Box-m^2)\phi(z)=0\,.
\end{align}
The above considerations are general, but we now turn to a specific graph, namely the Bruhat-Tits tree $T_{p^d}$ for the degree $d$ unramified extension of the $p$-adics, ie. the fractal tree of degree $p^d+1$. In this case, if we pick any vertex $z_0$ on the tree as a reference point, we can write down a basis of solutions $K_\Delta(z,x)$ parametrized by boundary points $x$ of the Bruhat-Tits tree. These solutions $K_\Delta(z,x)$ are identical to the bulk-to-boundary propagator of Ref.~\cite{Gubser:2016guj}, and they can be defined by the following two conditions:
\begin{enumerate}
\item The first condition is that each basis element equals unity at the reference point, 
\begin{align}
K_\Delta(z_0,x)=1\,. 
\end{align}
This condition amounts to a normalization convention.
\item Given any vertex $z$ on the tree, there is a unique adjacent vertex $z_x$ in the direction towards $x$. The second condition stipulates that
\begin{align}
K_\Delta(z_x,x)=p^{\Delta}K_\Delta(z,x)\,,
\end{align}
where $\Delta$ is a solution to the equation
\begin{align}
\label{mass}
m^2=p^{d-\Delta}+p^\Delta-p^d-1\,.
\end{align}
For $m^2>0$, equation \eqref{mass} has two solutions, one that is larger than $d$ and one that is smaller. We will use $\Delta$ to denote the larger solution.
\end{enumerate}
From the above equations, it as a simple exercise to check that
\begin{align}
(\Box-m^2)K_\Delta(z,x)=0\,.
\end{align}
So much for the Bruhat-Tits tree. We now turn to the Tate curve $T_{p^d}^{(w)}$.

The solutions to the free massive equation of motion can be determined from the solutions to the equation of motion on the Bruhat-Tits tree via the method of images. Again there exists a basis of solutions $K_\Delta^{(w)}(z,x)$ parametrized by boundary points $x$. In this case it is more convenient not to pick one global reference point, but to let the reference point depend on the boundary point. Specifically, for a given boundary point $x$, let $z_C(x)$ be the vertex on the thermal cycle in the Tate-curve that is closest to $x$. In the notation of Section~\ref{sec:Introduction}, $z_C(x)$ is the unique vertex on the thermal cycle for which $z_C(x) \rightarrow x$. We also need to introduce a type of bulk-to-boundary propagator $K(\gamma_{z,x})$ that depends not just on bulk point $z$ and boundary point $x$, but on the path $\gamma_{z,x}$ between them, and that is defined by the following conditions: 
\begin{itemize}
\item For the direct, shortest path between $x$ and the reference point $z_C(x)$, we assign $K(\gamma_{z,x})$ a value of one times the dimensionful quantity $|\beta|^{-\Delta}$. 
\item For every edge by which we extend a path, $K(\gamma_{z,x})$ picks up a factor of $p^{-\Delta}$. For paths that wind around the thermal circle, each edge contributes a factor of $p^{-\Delta}$ every time that same edge is traversed.
\item For every edge by which we shorten a path, $K(\gamma_{z,x})$ picks up a factor of $p^{\Delta}$. 
\end{itemize}
With the above definition in place, we can now describe the solution $K_\Delta^{(w)}(z,x)$. Given a boundary point $x$, the associated solution $K_\Delta^{(w)}(z,x)$ to the equation of motion is given by the sum of $K(\gamma_{z,x})$ over all geodesics $\gamma_{z,x}$ between $x$ and $z$, 
\begin{align}
K_\Delta^{(w)}(z,x) = \sum_{\text{geodesics }\gamma_{z,x}}K(\gamma_{z,x})\,,
\end{align}
where by ``geodesic", we mean a path between $x$ and $z$ that involves no back-tracking. This means that the sum over geodesics runs over the direct, shortest path plus the sum over any number of windings around the thermal cycle in the Tate curve in either direction. 

\begin{figure}
    \centering
\begin{align*}
\begin{matrix}\text{
\includegraphics[scale=0.8]{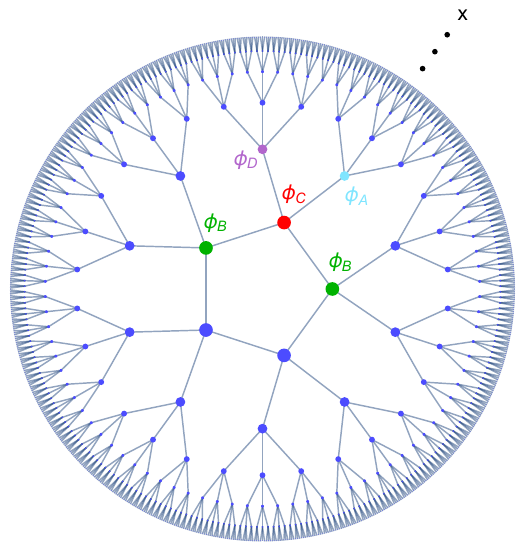}}
\end{matrix}
\end{align*}
    \caption{Field values on the Tate curve. For this example, $p^d=3$ and $w=5$.}
    \label{fig:example}
\end{figure}

To help clarify the above description, let us consider an example and verify that $K_\Delta^{(w)}(z,x)$ satisfies the equation of motion for the specific bulk point $z=z_C(x)$. As in Figure~\Ref{fig:example}, let $\phi_C$ denote the value of this solution at $z_C(x)$, let $\phi_A$ denote the value of the solution at the vertex point neighbouring $z_C$ in the direction of $x$, let $\phi_B$ denote the value of the solution at either of the two neighbouring vertices of $z_C$ that are also situated on the thermal cycle, and let $\phi_D$ denote the value of the solution on the remaining vertices adjacent to $z_C$. We have that
\begin{align}
\nonumber\\[-14mm]
\phi_C &= \frac{1}{|\beta|^\Delta}\Big(1+2\sum_{n=1}^\infty p^{-nw\Delta }\Big)\,,
\\
\phi_B &=\frac{1}{|\beta|^\Delta}\Big(p^{-\Delta}+p^{-(w-1)\Delta}\Big)\sum_{n=0}^\infty p^{-nw\Delta }\,,
\\
\phi_A &=\frac{1}{|\beta|^\Delta}\Big(p^\Delta +2p^{-\Delta}\sum_{n=1}^\infty p^{-nw\Delta }\Big)\,,
\\
\phi_D &=\frac{1}{|\beta|^\Delta}\,p^{-\Delta}\Big(1+2\sum_{n=1}^\infty p^{-nw\Delta }\Big)\,.
\end{align}
Using the above formulas, it is a straightforward exercise to verify that the equation of motion is satisfied:
\begin{align}
(\Box-m^2)\phi_C = 2\phi_B+\phi_A+(p^d-2)\phi_D-(p^d+1+m^2)\phi_C=0\,.
\end{align}

\begin{figure}
    \centering
\begin{align*}
\begin{matrix}
\\[-23pt]
\text{
\includegraphics[scale=0.8]{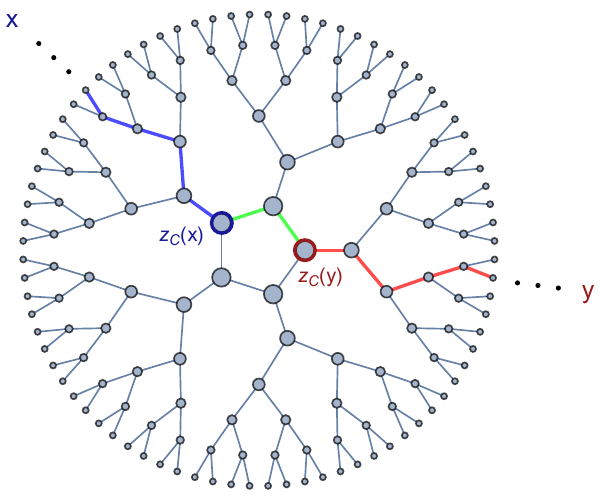}}
\end{matrix}
\hspace{13mm}
\begin{matrix}
\\[-19.5mm]
\text{
\includegraphics[scale=0.695]{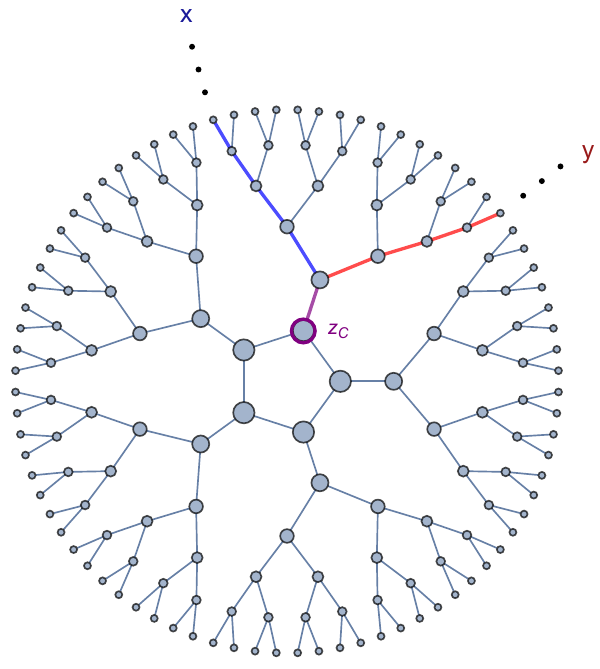}}
\end{matrix}
\end{align*}
    \caption{Left: The edge set $E_x$ is marked in blue, while the edges in $E_y$ are marked in red. The edge set $\gamma_{x,y}$ contains both these sets of edges and also the two green edges. Hence $H(\gamma_{x,y})=|\beta|^{-2\Delta}p^{-2\Delta}$. Right: The edge set $\gamma_{x,y}$ consists of the red and the blue edges. The multiset $E_{x,y}$ contains these same edges and also two copies of the purple edge. Hence $H(\gamma_{x,y})=|\beta|^{-2\Delta}p^{2\Delta}$.}
    \label{fig:BtoB}
\end{figure}

With a basis of solutions to the equation of motion in hand, we can proceed to solve the Dirichlet problem on the Tate curve. This problem consists in finding the unique solution to the equation of motion given suitable boundary conditions. For the class of solutions we are considering, each solution $K_\Delta^{(w)}(z,x)$ blows up as the bulk point $z$ approaches the boundary point $x$. To obtain a finite limit, we must perform a particular rescaling. Let $s(z,C)$ denote the number of edges separating the bulk point $z$ from the nearest bulk point on the thermal cycle. In Appendix~\ref{appendix:Dirichlet}, we derive the scaling that produces finite boundary fields $\widehat{\phi}(x)$ carrying suitable dimension:
\begin{align}
\label{scaling}
|\beta|^{\Delta-d}\lim_{z\rightarrow x}p^{-(\Delta-d)\,s(z,C)}\phi(z)=\widehat{\phi}(x)\,.
\end{align}
By defining $\epsilon(z) = |\beta|\,p^{-s(z,C)}$, it becomes apparent that \eqref{scaling} is the same kind of boundary scaling as in real AdS/CFT: 
 \begin{align}
 \label{fieldLimit}
\lim_{z\rightarrow x}\epsilon(z)^{\Delta-d}\phi(z)=\widehat{\phi}(x)\,.
\end{align}
The boundary conditions to the Dirichlet problem are stipulated by providing a complete list of boundary values $\widehat{\phi}(x)$. The solution to the Dirichlet problem is obtained by convolving the solution basis with the boundary conditions. This convolution is performed using a boundary integration measure that is defined via a limiting procedure of summing over all bulk vertices $z$ separated a distance $R$ from the thermal cycle in the limit as $R$ tends to infinity. More precisely, given a function $f(z)$ on the Tate curve that tends to a function $\widehat{f}(x)$ on the boundary as the bulk point $z$ approaches the boundary point $x$, the boundary integration measure is obtained from the following limit:
\begin{align}
\label{sumLimit}
|\beta|^d\lim_{R\rightarrow \infty }p^{-dR}\sum_{s(z,C)=R}f(z) = \int_{\partial T_{p^d}^{(w)}} dx\,\widehat{f}(x)\,.
\end{align}
With this definition in place, we are now in a position to state the solution to the Dirichlet problem on the Tate curve. Given a complete set of boundary values $\widehat{\phi}(x)$, the unique solution to the bulk equation of motion is given by 
\begin{align}
\label{DirichletSolution}
\phi(z)
=
\frac{\zeta_p(2\Delta)}{\zeta_p(2\Delta-d)}
\int_{\partial T_{p^d}^{(w)}} dy\,K_\Delta^{(w)}(z,y)\,\widehat{\phi}(y)\,.
\end{align}
We present an explicit derivation of this equation in Appendix~\ref{appendix:Dirichlet}, where we also derive another quantity that we will need in the next section, namely the boundary derivative for the Tate curve. By our boundary conditions, if we subtract from $\widehat{\phi}(x)$ the quantity $|\beta|^{\Delta-d}\,p^{-(\Delta-d)\,s(z,C)}\phi(z)$, then, in the limit as $z$ approaches $x$, the difference tends to zero. But by a second suitable rescaling, we obtain a finite limit for this difference. We define this finite limit as the boundary derivative at $x$. For the Tate curve, it turns out that the boundary derivative can conveniently be expressed in terms of a boundary-to-boundary propagator $H^{(w)}_\Delta(x,y)$:
\begin{align}
\label{boundaryDerivative}
\lim_{z\rightarrow x}
\Big(
\widehat{\phi}(x)
-\epsilon(z)^{\Delta-d}\,
\phi(z)\Big)
\epsilon(z)^{d-2\Delta}
=
\frac{\zeta_p(2\Delta)}{\zeta_p(2\Delta-d)}
\int_{\partial T_{p^d}^{(w)}} dy\,H^{(w)}_\Delta(x,y)\,\Big(\widehat{\phi}(x)
-\widehat{\phi}(y)\Big)\,.
\end{align}
The boundary-to-boundary propagator is defined in a similar manner to the way in which we defined the bulk-to-boundary propagator $K_\Delta^{(w)}(z,x)$. And so to define $H^{(w)}_\Delta(x,y)$, we first introduce a path-dependent boundary-to-boundary propagator $H(\gamma_{x,y})$ defined as follows:
\begin{itemize}
\item Given a boundary point $x$, let $E_x$ be the infinite set of edges connecting $x$ to $z_C(x)$. And given two boundary points $x$ and $y$, let $E_{x,y}$ be the multiset union of $E_x$ and $E_y$, meaning that duplicate elements are counted double (this can happen when $x$ and $y$ have the same valuation).
\item The edges of any path $\gamma_{x,y}$ can also be viewed as a multiset $E_{\gamma_{x,y}}$ since a path can traverse the same edge multiple times. The multisets $E_{x,y}$ and $E_{\gamma_{x,y}}$ differ by at most by a finite number of elements. Call this number $N_\text{diff}$.
\item  If we disallow paths that backtrack, then it will always be the case that either $E_{x,y} \subset E_{\gamma_{x,y}}$ or $E_{\gamma_{x,y}} \subset E_{x,y}$. If $E_{x,y} \subset E_{\gamma_{x,y}}$, we assign to $H^{(w)}_\Delta(x,y)$ a value of $|\beta|^{-2\Delta}p^{-\Delta N_\text{diff}}$, as in the left part of Figure~\ref{fig:BtoB}. If $E_{\gamma_{x,y}} \subset E_{x,y}$ (which implies that $z_C(x)=z_C(y)$), we assign to $H^{(w)}_\Delta(x,y)$ a value of $|\beta|^{-2\Delta}p^{\Delta N_\text{diff}}$, as in the right part of Figure~\ref{fig:BtoB}.
\end{itemize}
We then obtain $H^{(w)}_\Delta(x,y)$ from $H(\gamma_{x,y})$ by summing over all paths that do not involve backtracking:
\begin{align}
\label{Hwdef}
H^{(w)}_\Delta(x,y)=\sum_{\text{geodesics }\gamma_{x,y}}H(\gamma_{x,y})\,.
\end{align}
We explicitly evaluate this sum over geodesics in Section~\ref{subSec:tree}, which is devoted to studying the holographic two-point function at tree-level.

\section{Boundary Dual of the Massive Bulk Theory}
\label{sec:Boundary}

In 1988 Zabrodin derived the boundary action dual to the free massless bulk action on the Bruhat-Tits tree \cite{Zabrodin:1988ep}. At the time, this derivation was interpreted as relating the worldsheet action of the open $p$-adic string to the equivalent action on the boundary of the open worldsheet. After the discovery of the $p$-adic AdS/CFT, an alternative interpretation emerged wherein the discrete bulk tree-geometry is instead construed as the $p$-adic version of AdS space, while the $p$-adic numbers at the boundary of the tree are viewed as the Euclidean spacetime of a non-Archimedean CFT. Unlike the approach in Ref.~\cite{Zabrodin:1988ep}, where the boundary field configurations $\widehat{\phi}(x)$ were integrated over, the modern viewpoint identifies $\widehat{\phi}(x)$ with the source field for a CFT operator $\mathcal{O}$ and leaves the boundary field free so that the path integral becomes a functional of $\widehat{\phi}(x)$, which is related to the boundary CFT via the basic AdS/CFT identity
\begin{align}
\label{ads-cft}
Z_\text{bulk}[\widehat{\phi}]=\left<e^{i\int dx\,\widehat{\phi}(x)\,\mathcal{O}(x)}\right>_\text{CFT}\,.
\end{align}
In this section, adopting the viewpoint of $p$-adic AdS/CFT, we perform a two-fold generalization of Zabrodin's derivation: allowing for massive bulk fields and allowing for the bulk geometry to be given by the Tate curve with a thermal cycle of any length $w$. As we are presently dealing with a free theory, the bulk path becomes a quadratic functional of $\phi$. It is therefore possible to explicitly construct a quadratic action for $\mathcal{O}(x)$ such that Gaussian integration results in \eqref{ads-cft} being correctly reproduced, thereby deriving $p$-adic AdS/CFT in the trivial sense that a generalized free theory is a CFT. The subsequent sections proceed to introduce bulk interactions, which we can then study in perturbation theory around the integrable free theory.

Our starting point in this section is a regulated path integral $\mathcal{Z}_R$ that is only defined over a finite set of bulk points, namely those bulk point $z$ that are separated from the thermal cycle by at most $R$ edges. The regulated action depends on the field values over all these points, ie. on all $\phi(z)$ with $s(z,C) \leq R$, but in the path integral we only integrate over the fields with $s(z,C) < R$, so that the path integral becomes a function of boundary values $\phi_b(z)$, ie. the values of $\phi(z)$ with $s(z,C)=R$. At the end we take the $R\rightarrow\infty$ limit to arrive at a formula for the left-hand side of \eqref{ads-cft}. The regulated path integral is given explicitly by
\begin{align}
\mathcal{Z}_R[\phi_b]=\bigg[\int_{s(z,C)<R} D\phi(z)\bigg]\,e^{-S_R}\,,
\end{align}
where the regulated action $S_R$ decomposes into a bulk part and a boundary part:
\begin{align}
S_R = S_R^{\text{(bulk)}}+S_R^{\text{(boundary)}}\,,
\end{align}
with the two parts, allowing for an arbitrary overall normalization constant $\kappa$ in the action, given by
\begin{align}
S_R^{\text{(bulk)}} &= -\frac{\kappa}{2}\sum_{s(z,C)< R}\phi(z)(\Box-m^2)\phi(z)\,,
\\[5pt]
S_R^{\text{(boundary)}} &= -\frac{\kappa}{2}\sum_{s(z,C)= R}\phi(z)(\Box-M_b)\phi(z)\,.
\end{align}
We have introduced a modified mass-squared $M_b$ in the boundary piece. The reason for so doing is that in order to obtain a finite answer in the limit $R\rightarrow \infty$, it will turn out that  a special value of $M_b$ is required: 
\begin{align}
\label{msqb}
M_b = p^{d-\Delta}-1\,.
\end{align}
As this value is non-positive, we should not think of it as the square of a mass.

For any vertex $z$ not on the thermal circle, let $z_1$ be the unique neighbouring vertex in the direction of the thermal circle. In this notation, the boundary piece of the regulated action is given by
\begin{align}
S_R^{\text{(boundary)}} &= -\frac{\kappa}{2}\sum_{s(z,C)= R}\phi(z)\Big(\phi(z_1)-(1+M_b)\phi(z)\Big)\,.
\end{align}
Let us now decompose the bulk vertex field as $\phi(z)=\phi_0(z)+\phi_q(z)$, where $\phi_0(z)$ is equal to $\phi_b(z)$ when $s(z,C)=R$ and is uniquely specified for $s(z,C)<R$ by the requirement that it satisfy the equation of motion, $(\Box-m^2) \phi_0=0$, and where $\phi_q(z)$ can vary freely when $s(z,C)<0$ but must vanish when $s(z,C)=R$. Accordingly,
\begin{align}
\phi(z)
=\begin{cases}
\phi_0(z)+\phi_q(z)\hspace{10mm}\text{for }s(z,C)<R\,,
\\[5pt]
\phi_0(z)=\phi_b(z)\hspace{9.7mm}\text{for }s(z,C)=R\,.
\end{cases}
\end{align}
Under this decomposition, the regulated action can be written as
\begin{align}
S_R=\,& -\frac{\kappa}{2}\sum_{s(z,C)< R}\Big(\phi_0(z)+\phi_q(z)\Big)(\Box-m^2)\phi_q(z)
\nonumber\\[-9pt]
\\[-9pt] \nonumber
&-\frac{\kappa}{2}
\sum_{s(z,C)= R}\phi_0(z)\Big(\phi_0(z_1)+\phi_q(z_1)-(1+M_b)\phi_0(z)\Big)\,.
\end{align}
Applying summation by parts to the bulk cross term, the kinetic operator $(\Box-m^2)$ can be moved from $\phi_q$ to $\phi_0$ (where it gives zero) at the cost of a boundary term:
\begin{align}
\sum_{s(z,C)< R}\phi_0(z)(\Box-m^2)\phi_q(z)
=\,&
\sum_{s(z,C)< R}\phi_q(z)(\Box-m^2)\phi_0(z)
-
\sum_{s(z,C)= R}\phi_0(z)\phi_q(z_1)
\nonumber \\[-9pt]
\\[-9pt] \nonumber
=\,&
-
\sum_{s(z,C)= R}\phi_0(z)\phi_q(z_1)\,.
\end{align}
The regulated action can therefore be reexpressed as
\begin{align}
S_R= -\frac{\kappa}{2}\sum_{s(z,C)< R}\phi_q(z)(\Box-m^2)\phi_q(z)
-\frac{\kappa}{2}
\sum_{s(z,C)= R}\phi_0(z)\Big(\phi_0(z_1)-(1+M_b)\phi_0(z)\Big)\,.
\end{align}
We can get rid of unimportant overall factors in the path integral by choosing the normalization such that the path integral with vanishing boundary fields is set to one, which is tantamount to considering a functional $Z_R[\phi_b]$ obtained after division by $\mathcal{Z}_R[0]$:
\begin{align}
Z_R[\phi_b] = \frac{\mathcal{Z}_R[\phi_b]}{\mathcal{Z}_R[0]}\,.
\end{align}
On performing this normalization, the $\phi_q$ dependence drops out and we are left with the following:
\begin{align}
\label{A}
&\hspace{-8mm}Z_R[\phi_b] =
\exp\bigg\{
\frac{\kappa}{2}
\sum_{s(z,C)= R}\phi_0(z)\Big(\phi_0(z_1)-(1+M_b)\phi_0(z)\Big)
\bigg\}\,.
\end{align}
To massage \eqref{A} into a form where the $R\rightarrow \infty$ limit is more readily apparent, we first introduce the following shorthand:
\begin{align}
\label{Isum}
I=\sum_{s(z,C)= R}\phi_0(z)\Big(\phi_0(z_1)-(1+M_b)\phi_0(z)\Big)\,.
\end{align}
We can split the sum \eqref{Isum} into two parts,
\begin{align}
I = I_1+I_2\,,
\end{align}
where, subtracting off a piece from $I_1$ and adding back in an identical piece to $I_2$, the two parts are given by
\begin{align}
I_1&=\sum_{s(z,C)= R}\phi_0(z)\Big(\phi_0(z_1)-|\beta|^{d-\Delta}\,p^{(\Delta-d)(R-1)}\,\widehat{\phi}(x)\Big)\,,
\label{I1}
\\[5pt]
I_2&=-(1+M_b)\sum_{s(z,C)= R}\phi_0(z)\Big(\phi_0(z)-|\beta|^{d-\Delta}\,p^{(\Delta-d)R}\,\widehat{\phi}(x)\Big)\,.
\label{I2}
\end{align}
At this point it becomes clear why the boundary mass was chosen to have the specific value given in equation \eqref{msqb}: this precise choice effects a cancellation between the $\widehat{\phi}(x)$ terms introduced to \eqref{I1} and \eqref{I2},
\begin{align}
-p^{(\Delta-d)(R-1)}\widehat{\phi}(x)+(1+M_b)p^{(\Delta-d)R}\widehat{\phi}(x)=0\,.
\end{align}
The reason we added and subtracted the $\widehat{\phi}(x)$ terms was to bring each of $I_1$ and $I_2$ into a form where it has a finite limit as we let $R$ tend to infinity. To manifest this limit more clearly, we can regroup the factors in each case:
\begin{align}
I_1=\,&-p^\Delta\bigg(|\beta|^{d}\,p^{-dR}\sum_{s(z,C)= R}\bigg)\Big(|\beta|^{\Delta-d}p^{-(\Delta-d)R}\phi_0(z)\Big) \times
\\ \nonumber
&
\Big(\widehat{\phi}(x)-|\beta|^{\Delta-d}\,p^{-(\Delta-d)(R-1)}\,\phi_0(z_1)\Big)|\beta|^{d-2\Delta}\,p^{(2\Delta-d)(R-1)}\,,
\\[8pt]
I_2=\,&(1+M_b)\bigg(|\beta|^{d}\,p^{-dR}\sum_{s(z,C)= R}\bigg)\Big(|\beta|^{\Delta-d}p^{-(\Delta-d) R}\phi_0(z)\Big) \times
\\ \nonumber
&
\Big(\widehat{\phi}(x)-|\beta|^{\Delta-d}p^{-(\Delta-d)R}\,\phi_0(z)\Big)|\beta|^{d-2\Delta}\,p^{(2\Delta-d) R}\,.
\end{align}
Using the formula \eqref{boundaryDerivative} for the boundary derivative from the previous section as well as the limits \eqref{fieldLimit} and \eqref{sumLimit}, we see that the $R\rightarrow \infty$ limit yields the following:
\begin{align}
I_1&\rightarrow -p^{\Delta}\frac{\zeta_p(2\Delta)}{\zeta_p(2\Delta-d)}\int_{\partial T_{p^d}^{(w)}} dy\,dx\,\widehat{\phi}(x)\,H_\Delta^{(w)}(x,y)\,\Big(\widehat{\phi}(x)-\widehat{\phi}(y)\Big)\,,
\\[5pt]
I_2&\rightarrow (1+M_b)\frac{\zeta_p(2\Delta)}{\zeta_p(2\Delta-d)}\int_{\partial T_{p^d}^{(w)}} dy\,dx\,\widehat{\phi}(x)\,H_\Delta^{(w)}(x,y)\,\Big(\widehat{\phi}(x)-\widehat{\phi}(y)\Big)\,.
\end{align} 
Note that since $H_\Delta^{(w)}(x,y)$ is symmetric in its two arguments, we are allowed to perform the substitution
\begin{align}
\widehat{\phi}(x)\Big(\widehat{\phi}(x)-\widehat{\phi}(y)\Big)
\rightarrow \frac{1}{2}\Big(\widehat{\phi}(x)-\widehat{\phi}(y)\Big)^2\,.
\end{align}
To combine the coefficients when we add $I_1$ and $I_2$ together, we can use the identity
\begin{align}
-p^{\Delta}+(1+M_b)=-\frac{p^\Delta}{\zeta_p(2\Delta-d)}\,.
\end{align}
In summary, the normalized path integral over the entire Tate curve, $Z_\text{bulk}=Z_\infty$, equates to the following functional of the boundary field:
\begin{align}
\label{Zbulk}
Z_\text{bulk}[\widehat{\phi}]=
\exp\bigg\{
-\frac{\kappa\, p^\Delta \zeta_p(2\Delta)}{4\,\zeta_p(2\Delta-d)^2} \int_{\partial T_{p^d}^{(w)}} dx\,dy\,H_\Delta^{(w)}(x,y)\,\Big(
\widehat{\phi}(x)-\widehat{\phi}(y)\Big)^2
\bigg\}\,.
\end{align}
By equation \eqref{ads-cft}, the quadratic CFT action for $\mathcal{O}(x)$ is constructed from the inverse to the kernel for $\widehat{\phi}(x)$ in equation \eqref{Zbulk}. The two-point correlator for $\mathcal{O}(x)$, that is, the double functional derivative with respect to $\widehat{\phi}(x)$, is given by this inverse kernel. In effect, the present section has served to confirm that, up to a choice of normalization effected by the choice of $\kappa$, the CFT two-point function is given by the boundary-to-boundary propagator described in the previous section. This two-point function, at tree-level and at loop level in a theory with bulk interactions, is the subject of Section~\ref{sec:TwoPoint}.

\subsection{Non-Thermal Limit}

The derivation we have presented of the free boundary action, given by minus the exponent in equation \eqref{Zbulk}, applies also to the non-thermal case where the bulk geometry is given by the Bruhat-Tits tree. But it turns out the presence of a mass term in the bulk action introduces a subtle ambiguity in the definition of the non-thermal boundary action.

A straightforward way to obtain the flat space boundary action is by taking the $w\rightarrow \infty$ limit. We can choose to assign coordinates to the boundary points in this limit such that the boundary action becomes
\begin{align}
\label{limitAction}
S=\frac{\kappa\, p^\Delta \zeta_p(2\Delta)}{4\,\zeta_p(2\Delta-d)^2}\,|\beta|^{2d} \int_{\mathbb{Q}_{p^d}} d^\ast x
\int_{\mathbb{Q}_{p^d}} d^\ast y\,H_\Delta(x,y)\,\Big(
\widehat{\phi}(x)-\widehat{\phi}(y)\Big)^2\,.
\end{align}
Here we have re-expressed the integration measure in terms of the multiplicative Haar measure on the unramified extension $\mathbb{Q}_{p^d}$ of degree $d$ of the $p$-adic numbers. This measure is related to the additive measure $dx$ on $\mathbb{Q}_{p^d}$ by the equation
\begin{align}
\label{multiplicativeMeasure}
d^\ast x = \frac{dx}{|x|_p^d}\,,
\end{align}
where $|\cdot|_p$ denotes the $p$-adic norm on $\mathbb{Q}_{p^d}$. The kernel in the action in equation \eqref{limitAction} is the non-thermal boundary-to-boundary propagator, which is given by
\begin{align}
\label{limitH}
H_\Delta(x,y) =\,& \lim_{w\rightarrow \infty} H^{(w)}_\Delta(x)
= \frac{1}{|\beta|^{2\Delta}} \, \frac{|x|^\Delta_p\,|y|^\Delta_p}{|x-y|_p^{2\Delta}}\,.
\end{align}
To derive the RHS equation, first recall the definition of $H_\Delta^{(w)}(x,y)$ in \eqref{Hwdef}, and note that the $w\rightarrow \infty$ limit eliminates the contribution from any path that winds around the thermal cycle. For the sole remaining path $\gamma_{x,y}$, it is straightforward to check from the definition of $H_{\gamma_{x,y}}$ that when $|x|_p> |y|_p$, then $H_{\gamma_{x,y}}=|\beta|^{-2\Delta}\,|y|_p^\Delta/|x|_p^\Delta$, and when $|x|_p=|y|_p$, then $H_{\gamma_{x,y}}=|\beta|^{-2\Delta}\,(|x|_p/|x-y|_p)^{2\Delta}$, in accordance with \eqref{limitH}.

Plugging \eqref{multiplicativeMeasure} and \eqref{limitH} into \eqref{limitAction} yields a boundary action expressed in terms of the standard $p$-adic norm and integration measure. The factors of $|\beta|$ do not cancel but can be absorbed into $\kappa$ so that the thermal dependency disappears.

An alternative way of obtaining a boundary action for the free massive theory on the Bruhat-Tits tree follows Zabrodin's method more closely.\footnote{This method essentially consists in collapsing the thermal cycle into a single vertex $z_c$, except that so doing would result in $z_c$ having degree $w(p^d-1)$ rather than $p^d+1$.} In this formalism, one chooses a single bulk vertex as reference point $z_c$ and defines boundary fields and integration measure via a limiting procedure of moving away from this point. Given two boundary points $x$ and $y$, there will be a unique vertex $z'$ that is closest to $z_c$ among the vertices that lie on the direct bi-infinite path connecting $x$ and $y$ (this vertex $z'$ can possibly be $z_c$ itself). Let $n_c(x,y)$ denote the number of edges separating $z_c$ and $z'$. In that case, we can define a boundary distance function via
\begin{align}
\label{ZabrodinDistance}
|x,y|=p^{-n_c(x,y)}\,.
\end{align}
For any bulk point $z$, let $m_c(z)$ denote the number of edges separating $z$ and $z_c$. We define a boundary integration measure $d\mu_0(x)$ by demanding that, given a bulk function $f(z)$ that tends to a boundary function $\widetilde{f}(z)$ as $z$ approaches $x$, the following holds:
\begin{align}
\label{ZabrodinMeasure}
\int_{\mathbb{Q}_{p^d}} d\mu_0(x)\,\widetilde{f}(x) = \lim_{R\rightarrow \infty}p^{-dR}\sum_{m_c(z)=R}f(z)\,.
\end{align}
The integration measure obtained from this limit is a type of measure known in the mathematics literature as the Patterson-Sullivan measure \cite{Patterson:1976limit,Sullivan:1979density} and also studied in the physics literature in Ref.~\cite{Heydeman:2018qty}.

Boundary fields are obtained from bulk fields via the limiting procedure
\begin{align}
\label{phiLimitAlt}
\widetilde{\phi}(x)=\lim_{z\rightarrow x}p^{-(\Delta-d)m_c(z)}\phi(z)\,.
\end{align}
With these conventions in place, the resulting boundary action is given by
\begin{align}
S_\text{alt}=\frac{\kappa\, p^\Delta \zeta_p(2\Delta)}{4\,\zeta_p(2\Delta-d)^2} \int_{\partial T_{p^d}} \frac{d\mu_0(x)\,d\mu_0(y)}{|x,y|^{2\Delta}}\,\Big(
\widetilde{\phi}(x)-\widetilde{\phi}(y)\Big)^2\,.
\end{align}
To express this action in terms of the standard $p$-adic norm and measure, one can assign $p$-adic coordinates to the boundary points in such a way that $z_c$ is identified with the unique bulk vertex that sits at the intersection of the three geodesics connecting the numbers $x=0$, $x=1$, and $x=\infty$. In that case, as observed in Ref.~\cite{Zabrodin:1988ep}, the measure defined via \eqref{ZabrodinMeasure} is related to the additive $p$-adic Haar measure via
\begin{align}
d\mu_0(x)=\begin{cases}
    \hspace{1mm}dx\hspace{10mm}
    &\text{for }|x|_p\leq 1\,,
    \\[5pt]
    \displaystyle\frac{dx}{|x|_p^{2d}}
    &\text{for }|x|_p\geq 1\,,
\end{cases}
\end{align}
and the distance function \eqref{ZabrodinDistance} is related to the $p$-adic norm via
\begin{align}
|x,y|=\left\{\begin{array}{ll}
    |x-y|_p
    &\hspace{7mm}
    \text{for }|x|_p\leq 1\,,\hspace{2mm}
    |y|_p\leq 1\,,
    \\[5pt]
  \left|\frac{1}{x}-\frac{1}{y}\right|_p
    &\hspace{7mm}
    \text{for }|x|_p\geq 1\,,\hspace{2mm}
    |y|_p\geq 1\,,
        \\[5pt]
        1
        &\hspace{7mm}
        \text{otherwise}\,.
\end{array}\right.
\end{align}
The alternative boundary action can therefore be written as
\begin{align}
\label{alternativeBoundaryAction}
S_\text{alt}=\frac{\kappa\, p^\Delta \zeta_p(2\Delta)}{4\,\zeta_p(2\Delta-d)^2} 
\int_{\mathbb{Q}_{p^d}} dx 
\int_{\mathbb{Q}_{p^d}} dy\,\frac{|1,x|_s^{2\Delta-2d}\,|1,y|_s^{2\Delta-2d}}{|x-y|_p^{2\Delta}}\,\Big(
\widetilde{\phi}(x)-\widetilde{\phi}(y)\Big)^2\,,
\end{align}
where $|1,x|_s$ denotes the supremum of $1$ and $|x|_p$. The two boundary actions \eqref{limitAction} and \eqref{alternativeBoundaryAction} take on different guises because of the distinct manners in which the boundary fields were defined in equations \eqref{scaling} and \eqref{phiLimitAlt}. In the former case, boundary fields are scaled down from the bulk field depending on the distance to the thermal cycle, which in the $w\rightarrow \infty$ limit becomes a bi-infinite path through the Bruhat-Tits tree, whereas in the latter case the scaling is performed with respect to a single fixed boundary vertex. Under the coordinatization of the boundary points we performed in this section, the two types of boundary fields are related via
\begin{align}
\label{twoPhis}
\widetilde{\phi}(x) = \frac{1}{|\beta|^{\Delta-d}}\,
\widehat{\phi}(x)\times
\left\{
\begin{matrix}
|x|^{\Delta-d} \hspace{5mm}\text{for }|x|_p\leq 1
\\[4pt]
|x|^{d-\Delta}\hspace{5mm}\text{for }|x|_p\geq 1
\end{matrix}
\right\}
=
\frac{1}{|\beta|^{\Delta-d}}\,
\widehat{\phi}(x)\,
\frac{|x|^{\Delta-d}}{|1,x|_s^{2\Delta-2d}}\,.
\end{align}
The physically relevant parts of the two boundary actions are the cross term pieces containing $\widehat{\phi}(x)\widehat{\phi}(y)$ and $\widetilde{\phi}(x)\widetilde{\phi}(y)$ respectively, and these parts of the actions can be seen to be identical on invoking \eqref{twoPhis}. The diagonal $\widehat{\phi}(x)\widehat{\phi}(x)$ and $\widetilde{\phi}(x)\widetilde{\phi}(x)$ pieces are a form of regulator term and do no match for the two boundary actions, for the reason that the notion of well-behaved locally constant boundary functions are slightly different in the two formalisms.

In the massless case where $\Delta=d$, the two boundary actions become exactly identical, with the kernel being given simply by $|x-y|^{-2d}$. It is remarked in a footnote of Ref.~\cite{Gubser:2017tsi} that for a free $p$-adic field theory with scaling dimension $\Delta\in \mathbb{N}$, the cubic and quartic couplings of the bulk dual theory vanish at tree-level. We observe that this is an exact result. But the generalized free $p$-adic theory with non-integer scaling dimension is not dual to the free bulk theory.

\section{Two-Point Correlator}
\label{sec:TwoPoint}

Up to a choice of normalization convention, the free CFT two-point function on the boundary of the Tate curve is given by the boundary-to-boundary propagator:
\begin{align}
\left<\mathcal{O}_\Delta(x_1)\mathcal{O}_\Delta(x_2)\right>^{(0)}
=H_\Delta^{(w)}(x,y)
\,. 
\end{align}
This can be seen by twice performing a functional derivative of equation \eqref{ads-cft} with the left-hand side given in equation \eqref{Zbulk}. In the presence of bulk interactions, this two-point function is subject to corrections that can be evaluated perturbatively using Witten diagrams. The following two subsections \ref{subSec:tree} and \ref{subSec:loop} study the two-point function at tree level and one-loop level respectively.

\subsection{Tree level}
\label{subSec:tree}

In Section~\Ref{sec:Dirichlet}, we provided a prescription for evaluating the boundary-to-boundary propagator $H_\Delta^{(w)}(x,y)$. From that description it follows that $H_\Delta^{(w)}(x,y)$ assumes a rather different-looking functional form depending on whether or not $x$ and $y$ emanate from the same vertex on the thermal cycle, ie. whether or not $x$ and $y$ have the same valuation. As mentioned in Section~\ref{sec:Introduction}, the two cases are distinguished by whether or not $x$ and $y$ are so situated that the OPE is valid. We will consider them in turn and recover equations (6.2) and (6.3) of Ref.~\cite{Heydeman:2016ldy}.

First then, let us consider two boundary points $x_1$ and $x_2$ whose valuations are equal mod $w$, ie. $v(x_1)\equiv v(x_2)$. In this case the shortest bi-infinite path through the bulk connecting the two points will reach a vertex of closest approach to the thermal cycle. Let $l$ be the number of edges separating this bulk point from the vertex $z_C(x_1)=z_C(x_2)$. This number $l$ is related to the boundary norm we introduced in Section~\ref{sec:Introduction} via the relation
\begin{align}
\label{lrelation}
p^{-l}=\frac{|x_1-x_2|}{|\beta|}\,.
\end{align}
Explicitly carrying out the sum over all geodesics so as to evaluate the boundary-to-boundary propagator in accordance with the description given in Section~\Ref{sec:Dirichlet}, the two-point correlator evaluates to the following:
\begin{align}
\label{OOagain}
\left<\mathcal{O}_\Delta(x_1)\mathcal{O}_\Delta(x_2)\right>^{(0)}
=
\frac{1}{|x_1-x_2|^{2\Delta}}\bigg(
1+2p^{-2\Delta\,l} \sum_{n=1}^\infty p^{-nw\Delta}
\bigg)
=\frac{1}{|x_1-x_2|^{2\Delta}}
+\frac{2\zeta_p(w\Delta)p^{-w\Delta}}{|\beta|^{2\Delta}}\,,
\\[-44pt] \nonumber
\end{align}
resulting in the answer we previously mentioned in Section~\ref{sec:Introduction}.

This result is analogous to the Archimedean result on a cylinder, instead of a torus. The key is that both the bulk of the Tate curve and the cylinder deformation retracts to a circle.

Let us now consider the case when the boundary points have unequal valuations so that the OPE does not apply. Let $u=v(x_1)-v(x_2)$. By assumption $u$ is not divisible by $w$. The two-point correlator is given in this case by
\begin{align}
\label{OOoutside}
\left<\mathcal{O}_\Delta(x_1)\mathcal{O}_\Delta(x_2)\right>^{(0)}
&=\frac{1}{|\beta|^{2\Delta}}
\sum_{n\in\mathbb{Z}}p^{-\Delta|u+wn|}
\\ \nonumber
&=\frac{-1}{|\beta|^{2\Delta}}
\bigg(
\frac{2}{\log p^{w\Delta}}
+\frac{1}{\pi}
\text{Im}\Big[
L\big(\frac{u}{w},1,\frac{i\log p^{w\Delta}}{2\pi}\big)
+
L\big(-\frac{u}{w},1,\frac{i\log p^{w\Delta}}{2\pi}\big)
\Big]
\bigg)
\,,
\end{align}
where $L(\lambda,s,\alpha)$ is the Hurwitz-Lerch zeta function. We observe that the right-hand side is periodic in $u$ with periodicity $w$, in accordance with the geometry of the problem. To obtain a simpler form for the correlator, we can choose for $u$ to lie inside the fundamental domain $0 < u <w$, in which case the sum in \eqref{OOoutside} evaluates to the following:
\begin{align}
\left<\mathcal{O}_\Delta(x_1)\mathcal{O}_\Delta(x_2)\right>^{(0)}
&=\zeta_p(w\Delta)\frac{p^{-\Delta u}+p^{-\Delta(w-u)}}{|\beta|^{2\Delta}}
 \nonumber \\[-8pt]
 \label{uinrange}
\\[-8pt] \nonumber
&=\frac{2\zeta_p(w\Delta) p^{-\Delta w/2}}{|\beta|^{2\Delta}}
\cosh\Big(\log p \,\Delta (u-w/2)\Big)\,.
\end{align}

\subsubsection{$\Delta \rightarrow 0$ limit}

Let us consider the limit of vanishing $\Delta$. This limit commands interest because it generates logarithmic propagators. In two-dimensional real theories, such a propagator follows from a standard two-derivative kinetic term, but the scalar requires a non-compact spacetime. Expanding equations \eqref{OOagain} and \eqref{uinrange} in $\Delta$ gives
\begin{align}
\label{deltaExpand1}
&\hspace{47mm}\left<\mathcal{O}_\Delta(x_1)\mathcal{O}_\Delta(x_2)\right>^{(0)}\Big|_{u=0}
=\,
\\[2mm] \nonumber
&\frac{2}{\Delta\log p^w}
-\frac{4\log |\beta|}{\log p^w}
+
\bigg(
\frac{4(\log |\beta|)^2}{\log p^w}
+\frac{1}{6}\log p^w
+\log \frac{|\beta|^2}{|x_{1,2}|^2}
\bigg)\Delta
+\mathcal{O}(\Delta^2)\,,
\\[5mm]
\label{deltaExpand2}
&\hspace{47mm}\left<\mathcal{O}_\Delta(x_1)\mathcal{O}_\Delta(x_2)\right>^{(0)}\Big|_{0<u<w}
=\,
\\[2mm] \nonumber
&\frac{2}{\Delta\log p^w}
-\frac{4\log |\beta|}{\log p^w}
+
\bigg(
\frac{4(\log |\beta|)^2}{\log p^w}
+\frac{1}{6}\log p^w
-u \log p
+\frac{u^2}{w}\log p 
\bigg)\Delta
+\mathcal{O}(\Delta^2)\,.
\end{align}
In order to get a finite logarithmic piece in the $\Delta\rightarrow 0$ limit, we need to change the overall normalization of the correlator so that it carries an extra factor of $1/\Delta$. This can be achieved for example by introducing a factor $\zeta_p(2\Delta)$, which in fact was the normalization chosen in \cite{Jepsen:2018dqp} on the grounds that it simplified Mellin space formulas. 

After changing normalization, the propagator contains $1/\Delta^2$ and $1/\Delta$ terms, but these can be dealt with by a suitable constant shift of the operators, $\mathcal{O}_\Delta \rightarrow \mathcal{O}_\Delta + c$. The cross term gives a thermal one-point function that has the expansion
\begin{align}
\left<\mathcal{O}_\Delta\right> = \frac{b}{|\beta|^\Delta}
=b - b \Delta\log |\beta|
+\frac{b\Delta^2}{2}(\log |\beta|)^2
+\mathcal{O}(\Delta^3)\,,
\end{align}
and provided the one-point coefficient $b$ assumes the appropriate value, the $\beta$-dependent subleading terms in \eqref{deltaExpand1} and \eqref{deltaExpand2} are then cancelled. Crucially, this way of obtaining finite propagators in the $\Delta \rightarrow 0$ limit is possible because the leading and subleading terms in \eqref{deltaExpand1} and \eqref{deltaExpand2} are identical.

For the resultant logarithmic correlator, we observe that the leading terms of the OPE and non-OPE region correlators also match in the $w\rightarrow \infty$ limit, in the sense that their ratio tends to one (their difference does not tend to zero). This behavior is not affected by any possible normalization which may bring in corrections from the leading and sub-leading terms, as such corrections would be of lower order in $w$. The same observation does not hold when $\Delta>0$, where the $w \rightarrow \infty$ correlator is independent of $\beta$ in the OPE region but depends on $\beta$ outside the OPE region.

If one fixes $w$ but takes the $|\beta| \rightarrow \infty$ limit of the logarithmic correlator at $\Delta = 0$, then the leading terms again match for the two regions. Again this behavior is not affected by corrections coming from the pole terms because such corrections would be of lower order in $\log |\beta|$.

\subsection{One-loop level}
\label{subSec:loop}

In the presence of bulk interactions, the two-point correlators of the previous subsection are subject to corrections, which in the limit of weak couplings organize themselves into a perturbative expansion describable by Witten diagrams. The present subsection studies the leading perturbative correction induced by a quartic interaction in the bulk. The specific interaction we shall introduce is given as follows
\begin{align}
\label{quarticInteraction}
\sum_{z\in T_{p^d}^{(w)}} \
\phi_{\Delta_1}(z)\,
\phi_{\Delta_2}(z)\,
\phi_{\Delta_a}(z)^2\,,
\end{align}
where the subscripts on the fields indicate their scaling dimensions as determined by their respective masses. We are allowing fields with different masses to obtain a more general answer and to better exhibit the functional form of the distinct terms that appear in the OPE. We have taken two of the fields in the interaction term to carry the same scaling dimension $\Delta_a$ since otherwise the one-loop correction will be absent from the theory.
We empasize that the loop calculation of the present section is carried out in the discrete Tate-curve geometry, for which UV divergences are absent. There exists an alternative and distinct bulk formulation known as $p$AdS \cite{Gubser:2016guj}, in which loop computations receive contributions from degrees of freedom at smaller spacings than the edge lenght of the Bruhat-Tits tree \cite{Qu:2018ned}, and this formulation may provide a more faithful model for loop corrections in real AdS/CFT.

Unlike their non-thermal counterparts, thermal CFTs can have non-zero two-point functions for distinct operators. This is not apparent at tree-level, but at one-loop level such two-point functions pick up non-vanishing contributions. For the specific interaction we introduced in equation \eqref{quarticInteraction}, the one-loop diagram is given up to a constant prefactor by the following bulk integral:
\begin{align}
\label{D2int}
D_2^{(\text{1-loop})}
=\sum_{z\in T_{p^d}^{(w)}}
K_{\Delta_1}(x_1,z)\,
K_{\Delta_2}(x_2,z)\,
G_{\Delta_a}(z,z)\,,
\end{align}
where $K_\Delta(x_i,z)$ is the bulk-to-boundary correlator introduced in Section~\ref{sec:Dirichlet}. $G_\Delta(z,z')$ is the bulk-to-bulk propagator, obtained in the thermal case from the non-thermal propagator by performing a sum over thermal images. Concretely, this means that
\begin{align}
G_\Delta(z,z')=\sum_{\text{geodesics } \gamma_{z,z'}}G_\Delta(\gamma_{z,z'})\,,
\end{align}
where $G_\Delta(\gamma_{z,z'})$ is given by $p^{-\Delta |\gamma_{z,z'}|}$ with $|\gamma_{z,z'}|$ denoting the number of edges (counted with multiplicity) in the path $\gamma_{z,z'}$. 

The bulk-to-bulk propagator in \eqref{D2int} is evaluated at coincident points. At the real place, coincident points induce a divergence in the bulk-to-bulk propagator, which is set to zero by mass renormalization so that the only remaining contributions are from paths connecting distinct thermal images. At the $p$-adic place, however, the bulk-to-bulk propagator remains finite at coincident points since $G_\Delta(\gamma_{z,z})=1$ when $\gamma_{z,z}$ is the trivial path with no edges. But this contribution induces a separate divergence since the sum $\sum_{z\in T_{p^d}^{(w)}}
K_{\Delta_1}(x_1,z)\,
K_{\Delta_2}(x_2,z)$ does not converge. The divergence, however, takes the form of a geometric series, which can be formally evaluated, resulting in the equation
\begin{align}
\label{KKzero}
\sum_{z\in T_{p^d}^{(w)}}
K_{\Delta_1}(x_1,z)\,
K_{\Delta_2}(x_2,z)=0\,.
\end{align}
In brief, on allowing geometric summation irrespectively of convergence, the one-loop diagram \eqref{D2int} is insensitive to whether or not we include the trivial paths in the bulk-to-bulk propagator.

As at tree-level, the form of the one-loop correlator depends on whether or not the two boundary points $x_1$ and $x_2$ are so situated that the OPE applies. We will first consider the case when it does, and so we set $v(x_1)=v(x_2)$. In Appendix~\ref{appendix:twoPoint}, we explicitly evaluate the integral \eqref{D2int} over the Tate curve for such a boundary configuration. To display the answer, it is convenient to introduce the following shorthands:
\begin{align}
\Delta_{ijk} &\equiv \Delta_i+\Delta_j+\Delta_k\,,
\nonumber 
\\ \nonumber 
\Delta_{ij} &\equiv \Delta_i+\Delta_j\,,
\\ 
\label{shorthands}
\Delta_{ij,k} &\equiv \Delta_i+\Delta_j-\Delta_k\,,
\\ \nonumber 
\Delta_{ij,kl} &\equiv \Delta_i+\Delta_j-\Delta_k-\Delta_l\,,
\\ \nonumber 
x_{i,j} &\equiv x_i - x_j\,.
\end{align}
and also to introduce the following definition:
\begin{align}
\label{f}
f(\Delta_a|\Delta_b|\Delta_c) = \frac{
\zeta_p(\Delta_{abc}-d)\,\zeta_p(\Delta_{ab,c})\,\zeta_p(\Delta_{ac,b})\,\zeta_p(\Delta_{bc,a})}{\zeta_p(2\Delta_a)\,\zeta_p(2\Delta_b)\,\zeta_p(2\Delta_c)}\,,
\end{align}
where the function $\zeta_p(s)$ was defined in equation \eqref{OO}. Note that the function $f(\Delta_a|\Delta_b|\Delta_c)$ is precisely equal to the prefactor of the generic (non-thermal) CFT three-point function in $p$-adic AdS/CFT, as first determined in Ref.~\cite{Gubser:2017tsi}. The answer we arrive at for the one-loop Witten diagram for two boundary points in the OPE regime is given by
\begin{align}
\label{D2oneloop}
D_2^{(\text{1-loop,OPE})}
=
\bigg(&
\frac{f(\Delta_{12}|\Delta_a|\Delta_a)}{|\beta|^{\Delta_{12}}}+\frac{f(\Delta_{1a}|\Delta_2|\Delta_a)}{|\beta|^{\Delta_{12}}}\frac{2\zeta_p(w\Delta_1)}{p^{w\Delta_1}}+\frac{f(\Delta_1|\Delta_{2a}|\Delta_a)}{|\beta|^{\Delta_{12}}}\frac{2\zeta_p(w\Delta_2)}{p^{w\Delta_2}}
\nonumber
\\
&
+
\frac{f(\Delta_1|\Delta_2|2\Delta_a)}{|x_{1,2}|^{\Delta_{12,aa}}|\beta|^{2\Delta_a}}
\bigg)\frac{2\zeta_p(w\Delta_a)}{p^{w\Delta_a}}\,.
\end{align}

Let us next consider the one-loop diagram when the OPE does not converge, ie. when $v(x_1) \neq v(x_2)$. Without loss of generality, we can choose for $u=v(x_1) - v(x_2)$ to lie in the range $0<u<w$. In this case, as we show explicitly in Appendix~\ref{appendix:twoPoint}, evaluating the bulk integral produces the answer
\begin{align}
\label{D2oneloopAlt}
D_2^{(\text{1-loop,non-OPE})}
=
\frac{1}{|\beta|^{\Delta_{12}}}
\bigg(
&f(\Delta_{1a}|\Delta_2|\Delta_a)
\zeta_p(w\Delta_1)
\Big(p^{-\Delta_1 u}+p^{-\Delta_1(w-u)}\Big)
+
\\ \nonumber
&f(\Delta_{2a}|\Delta_1|\Delta_a)
\zeta_p(w\Delta_2)
\Big(p^{-\Delta_2 u}+p^{-\Delta_2(w-u)}\Big)
\bigg)
2p^{-w\Delta_a}\zeta_p(w\Delta_a)\,.
\end{align}

\subsubsection{Contour formula}
\begin{figure}
    \centering
\begin{align*}
\begin{matrix}\text{
\includegraphics[scale=0.8]{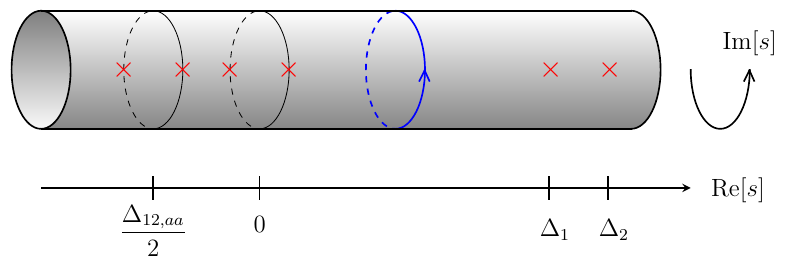}}
\end{matrix}
\end{align*}
    \caption{Cylindrical contour integral for the one-loop two-point Witten diagram.}
    \label{fig:cyl}
\end{figure}

In real thermal AdS/CFT, the analogous Witten diagram to the one-loop two-point diagram considered in this section was computed in Ref.~\cite{Alday:2020eua}, which found that their answer could be most elegantly expressed as a contour integral. For the $p$-adic answer, a similar contour formula also exists. Moreover, the formula enjoys the property of assuming the same form irrespectively of whether or not the boundary points are situated inside the OPE region:
\begin{align}
\label{OO1general}
D_2^{(\text{1-loop})}
=\,&
2p^{-w\Delta_a}\zeta_p(w\Delta_a)\,\zeta_p(\Delta_{12aa}-d)
\,\frac{\log p}{2\pi i}\,
\times
\\
&\hspace{-2mm}
\int_{c-\frac{i\pi}{\log p}}^{c+\frac{i\pi}{\log p}}ds\,
\frac{\zeta_p(\Delta_1-s)\zeta_p(\Delta_2-s)}{\zeta_p(2\Delta_1)\zeta_p(2\Delta_2)}\,
\frac{\zeta_p(\Delta_{12})\zeta_p(2s)\zeta_p(\Delta_{aa,12}+2s)}{\zeta_p(2\Delta_a+2s)\zeta_p(2\Delta_a)}\,
\frac{\left<\mathcal{O}_s(x_1)\mathcal{O}_s(x_2)\right>^{(0)}}{|\beta|^{\Delta_{12}-2s}}\,,
\nonumber
\end{align}
where the zeroth-order correlator $\left<\mathcal{O}_s(x_1)\mathcal{O}_s(x_2)\right>^{(0)}$ was given in equations~\eqref{OOagain} and \eqref{OOoutside}, and where $c$ is any complex number satisfying the conditions $\text{max}(0,\frac{\Delta_{12,aa}}{2})<\text{Re}[c]<\text{min}(\Delta_1,\Delta_2)$. Since the integrand is periodic in the imaginary direction, the integration contour wraps around a cylinder as shown in Figure~\ref{fig:cyl}. 

In the non-OPE case when $v(x_1)\neq v(x_2)$, the contour can be shifted to plus infinity where it vanishes, picking up in the process a residue at $s=\Delta_1$ and another at $s=\Delta_2$. The contributions from these two residues match the two terms in \eqref{D2oneloopAlt}.

In the OPE case when $v(x_2)=v(x_1)$, the contour integral can be evaluated separately for each of the two terms in $\left<\mathcal{O}_s(x_1)\mathcal{O}_s(x_2)\right>^{(0)}$ as given by the right-hand side of equation \eqref{OOagain}. For the $|x_{1,2}|^{-2s}$ term in $\left<\mathcal{O}_s(x_1)\mathcal{O}_s(x_2)\right>^{(0)}$, the contour can be shifted to Re$[s]=-\infty$, where it goes away, at the cost of four residues: two at $s=0$ and at $s=\frac{i \pi}{\log p}$, which sum up to give the first term in the right-hand side of equation \eqref{D2oneloop}, and two at $s=\frac{\Delta_{12,aa}}{2}$ and at $s=\frac{\Delta_{12,aa}}{2}+\frac{i \pi}{\log p}$, which sum up to give the last term in \eqref{D2oneloop}. For the $2\zeta_p(s\Delta)p^{-ws}|\beta|^{-2s}$ term, the contour can instead be shifted to Re$[s]=\infty$ where the integral vanishes, picking up along the way residues at $s=\Delta_1$ and $s=\Delta_2$, which respectively give the second and third terms in the right-hand side of equation \eqref{D2oneloop}.

Setting the three scaling dimensions in equation \eqref{OO1general} equal to $\Delta$ and dressing the equation with a symmetry factor of $\frac{1}{2}$ and a factor of minus one from the negative sign in the exponent of the Euclidean path integral, we recover equation \eqref{OO1} of the introduction.

\section{Three-Point Correlator}
\label{sec:ThreePoint}

In the present section we turn on generic weak cubic bulk interactions and present the holographic three-point correlator to leading perturbative order. Up to a constant prefactor, this correlator can be computed from the following bulk integral
\begin{align}
\label{OOO}
\left<
\mathcal{O}_{\Delta_1}(x_1)
\mathcal{O}_{\Delta_2}(x_2)
\mathcal{O}_{\Delta_3}(x_3)
\right>
=\sum_{z\in T_{p^d}^{(w)}}
K_{\Delta_1}(x_1,z)
K_{\Delta_2}(x_2,z)
K_{\Delta_3}(x_3,z)\,,
\end{align}
where $K_\Delta(x_i,z)$ is the bulk-to-boundary correlator introduced in Section~\ref{sec:Dirichlet}. To gain information about the OPE, we will consider a configuration of the boundary points for which the OPE is convergent. We consider boundary points, therefore, for which $v(x_1)=v(x_2)=v(x_3)$. Even among such configurations, the form of the position space correlator does not assume a universal form for the reason that there are special cases where the answer looks different, namely the cases when either $|x_1|$ or $|x_2|$ or $|x_3|$ is equal to $|\beta|$ or the cases when the norms $|x_1-x_2|$, $|x_1-x_3|$, and $|x_2-x_3|$ are all equal (by the ultrametric property of $p$-adics, there will always be two of these norms that are equal). We will present the correlator, not for these special cases, but for the generic configuration of boundary points in the OPE regime. 

In Appendix~\ref{appendix:threePoint} we explicitly perform the sum over all bulk points in \eqref{OOO} and also sum, for each propagator, over any number of windings around the thermal cycle, thereby determining the three-point correlator for the specified configuration of boundary points. Adopting the notation of \eqref{shorthands}, the three-point correlator evaluates to the following:
\begin{align}
\label{OOOanswer}
&\hspace{27mm}
\left<
\mathcal{O}_{\Delta_1}(x_1)
\mathcal{O}_{\Delta_2}(x_2)
\mathcal{O}_{\Delta_3}(x_3)
\right>
=
f(\Delta_1|\Delta_2|\Delta_3)\,
\times
\\[7pt] \nonumber
&\bigg(
\frac{1}{
|x_{1,2}|^{\Delta_{12,3}}
|x_{1,3}|^{\Delta_{13,2}}
|x_{2,3}|^{\Delta_{23,1}}}
+\frac{2\zeta_p(w\Delta_1)p^{-w\Delta_1}}{|x_{2,3}|^{\Delta_{23,1}}|\beta|^{2\Delta_1}}
+\frac{2\zeta_p(w\Delta_2)p^{-w\Delta_2}}{|x_{1,3}|^{\Delta_{13,2}}|\beta|^{2\Delta_2}}
+\frac{2\zeta_p(w\Delta_3)p^{-w\Delta_3}}{|x_{1,2}|^{\Delta_{12,3}}|\beta|^{2\Delta_3}}
\\[7pt] \nonumber
&
\hspace{4mm}
+
\frac{2\zeta_p(w\Delta_2)
\zeta_p(w\Delta_3)
p^{-w\Delta_{23}}}{|\beta|^{\Delta_{123}}}
+
\frac{2\zeta_p(w\Delta_1)
\zeta_p(w\Delta_3)
p^{-w\Delta_{13}}}{|\beta|^{\Delta_{123}}}
+
\frac{2\zeta_p(w\Delta_1)
\zeta_p(w\Delta_2)
p^{-w\Delta_{12}}}{|\beta|^{\Delta_{123}}}
\bigg)\,,
\end{align}
where the function $f(\Delta_1|\Delta_2|\Delta_3)$ was defined in equation \eqref{f}. The seven terms on the right-hand side of \eqref{OOOanswer} may be interpreted as the thermal three-point conformal blocks, which thus consist of a few elementary functions. Performing initial forays into the analogous real computation, we find that extending the thermal mean field two-point conformal block decomposition of Ref. \cite{iliesiu2018conformal} to three points in the most direct manner will involve nested infinite sums and the multiple zeta function.

\section{Outlook}
\label{sec:Outlook}

Through the computations presented in this paper, we hope to have convincingly established that the $p$-adic AdS/CFT correspondence extends farther than previously known, as finite-temperature CFTs too fall within the range of phenomena that are encompassed by the $p$-adic formalism. 

Compared to the technically quite challenging computations over the reals, $p$-adic correlators can be computed with relative facility, thereby allowing for the $p$-adic formalism to serve as a toy model to provide helpful instructions for real holography. We have here presented a few correlators for which the real counterparts have not yet been determined (namely for the one-loop two-point correlator at generic scaling dimension and for the tree-level three-point correlator), and there are no obstacles to proceeding to compute many more Witten diagrams in this setting. An eventual intermediate goal will be to write down Feynman rules for the thermal Witten diagrams in $p$-adic AdS/CFT. Once accomplished, this feat can serve to inspire and illuminate the more momentous challenge of determining those same Feynman rules at the real place. 

On the more mathematical side, the investigations presented in this paper also suggest some possible avenues for delving deeper into the structure of $p$-adic AdS/CFT and its relation to the real version. In many instances of flat space AdS/CFT, one can translate between real and $p$-adic answers by an interchange of local zeta functions $\zeta_p \leftrightarrow \zeta_\infty$. But a comparison of the real and $p$-adic two-point correlators in equations~\eqref{OOreal} and \eqref{OO} reveals that, in a black hole background, this translation is replaced with a local-global interchange of zeta functions: $\zeta_p \leftrightarrow \zeta$. It would be desirable to understand the relations between the special functions appearing in the two types of holography. 

A property of the $p$-adic thermal correlators that distinctly sets them apart from the real counterparts is their non-trivial dependence on a discrete parameter $w$. Real thermal CFT correlators do not carry such parameters, and so from a purely boundary perspective it is tempting to associate different values of $w$ to different thermal states and therefore also to different black hole geometries. But on the other hand the references cited in subsection~\ref{subSec:thermalFinite} have demonstrated that entanglement entropy on the Tate curve scales with $w$, which provides a compelling reason to associate this variable to a black hole radius. Future work is called for to show how this apparent tension may be resolved.

On the real side, the BTZ black hole is part of a larger family of solutions to the Einstein equations known as the $SL(2,\mathbb{Z})$ black holes. The sum over these geometries produces the gravitational path integral, and a series of beautiful papers, including \cite{Witten:2007kt,Maloney:2007ud,Maloney:2009ck}, have brought to light the many connections between this object and wormholes, CFT ensembles, modular transformations, and the $j$-invariant.  At present it is not know whether an action exists by which to perform a kind of gravitational path integral over discrete geometries, but if such an action may be found, then carefully performing the sum over $p$-adic geometries will reveal if more such beautiful stories remain to be told.

\subsection*{Acknowledgements}

We are grateful to Luca Iliesiu for sharing important insights. The work of A.~H. is supported by the Simons collaboration grant 708790. The work of C.~B.~J. is supported by the Korea Institute for Advanced Study (KIAS) Grant PG095901.

\appendix

\section{Explicit Computations}
\label{Appendix}

This appendix provides explicit computational details that were omitted from the main part of the paper. Subsection \ref{appendix:Dirichlet} provides a derivation of the solution to the Dirichlet problem on the Tate curve and the formula for the boundary derivative; subsection  \ref{appendix:twoPoint} carries out the computation of the one-loop two-point Witten diagram; and subsection \ref{appendix:threePoint} contains the calculation of the thermal $p$-adic three-point function at tree-level.

\subsection{Dirichlet problem and boundary derivative}
\label{appendix:Dirichlet}

In this section we derive the solution \eqref{DirichletSolution} to the Dirichlet problem on the Tate curve and the formula \eqref{boundaryDerivative} for the boundary derivative.

\begin{figure}
    \centering
\begin{align*}
\begin{matrix}\text{
\includegraphics[scale=1.3]{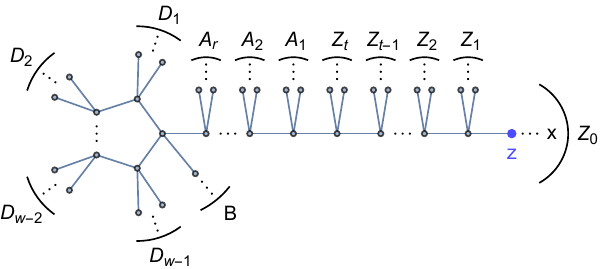}}
\end{matrix}
\end{align*}
    \caption{Partition of the boundary points $\partial T_{p^d}^{(w)}$ of the Tate curve into various subsets.}
    \label{fig:boundaryPartition}
\end{figure}

As mentioned in Section~\ref{sec:Dirichlet}, for any boundary point $x$ the bulk-to-boundary propagator $K^{(w)}_\Delta(x,z)$ furnishes a solution to the bulk equation of motion, as follows from the method of images. For rather than thinking of the Tate curve as a graph containing a finite thermal cycle at the center, we can think of the geometry as obtained by imposing periodicity along a bi-infinite path of the Bruhat-Tits tree, so that it follows that solutions to the equation of motion on the Tate curve are those linear combinations of solutions for the Bruhat-Tits tree that respect this periodicity. Since $K^{(w)}_\Delta(x,z)$ solves the equation of motion, so too does any linear combination of this function, in particular so does \eqref{DirichletSolution}. All that remains to be shown in order to establish that this equation gives the solution to the Dirichlet problem is that the correct boundary conditions \eqref{scaling} are satisfied. Consider now the integral
\begin{align}
\label{consideredIntegral}
\mathcal{I} = 
\int_{\partial T_{p^d}^{(w)}} dy\,K^{(w)}_\Delta(z,y)\,\widehat{\phi}(y)\,,
\end{align}
where $\widehat{\phi}(y)$ is some given function that encodes the boundary conditions to the problem. The type of well-behaved functions we allow for this boundary value problem are the so-called Schwartz-Bruhat functions, which in this context are simply the set of locally constant functions. This entails that if we let a bulk point $z$ tend toward a fixed boundary point $x$, eventually it will hold true that $\widehat{\phi}(y)=\widehat{\phi}(x)$ for every boundary point $y \in Z_0$ emanating from $z$ in the direction away from the thermal circle, as shown in Figure~\ref{fig:2ptDiagram}. On this figure, we have also labelled several other subsets of boundary points: $Z_k$ with $k\in\{1,...,t\}$, $A_j$ with $j\in\{1,...,r\}$, $D_i$ with $i\in\{1,...,w-1\}$, and $B$. Together, these subsets form a partition of the total set of boundary points of the Tate curve:
\begin{align}
\label{partition}
\partial T_{p^d}^{(w)}
=
B
\bigcup\limits_{i=1}^{w-1} D_i\,
\bigcup\limits_{j=1}^r A_j\,
\bigcup\limits_{k=0}^t Z_k\,.
\end{align}
The subsets $Z_k$ and $A_j$ are distinguished by the condition that we demand that $\widehat{\phi}(y)=\widehat{\phi}(x)$ for any $y\in Z_k$. For a given set of boundary conditions, if we take the boundary point $x$ to be fixed and the bulk point $z$ to be variable, then the integer $r$ will also be fixed ($\widehat\phi$ is locally constant, so it assumes a constant value on some ball near $x$, and $r$ is determined by the radius of this ball), whereas the integer $t$ will increase as we let $z$ move closer to $x$.

Let us now split up the integral \eqref{consideredIntegral} according to the partition \eqref{partition}. From the definition of the bulk-to-boundary propagator we gave in Section~\ref{sec:Dirichlet}, it is a simple exercise to show the following:
\begin{align}
\nonumber
\int_{B} dy\,K^{(w)}_\Delta(z,y)\,\widehat{\phi}(y)=
\,& \frac{1}{|\beta|^\Delta}\,p^{\Delta(-r-t-1)}
\bigg(1+2
\sum_{n=1}^\infty p^{-\Delta w n}
\bigg)
\int_{B} dy\,\widehat{\phi}(y)\,,
\\[5pt] \nonumber
\int_{D_i} dy\,K^{(w)}_\Delta(z,y)\,\widehat{\phi}(y)=\,& \frac{1}{|\beta|^\Delta}\,
p^{\Delta(-r-t-1)}\Big(p^{-\Delta i}+p^{\Delta(i-w)}\Big)\sum_{n=0}^\infty p^{-\Delta w n}
\int_{D_i} dy\,\widehat{\phi}(y)\,,
\\[-8pt]
\label{intPieces}
\\[-8pt] \nonumber
\int_{A_j} dy\,K^{(w)}_\Delta(z,y)\,\widehat{\phi}(y)=\,& \frac{1}{|\beta|^\Delta}\,
\bigg(p^{\Delta(r+1-t-2j)}+2p^{\Delta(-r-t-1)}
\sum_{n=1}^\infty p^{-\Delta w n}
\bigg)
\int_{A_j} dy\,\widehat{\phi}(y)\,,
\\[5pt] \nonumber
\int_{Z_k} dy\,K^{(w)}_\Delta(z,y)\,\widehat{\phi}(y)=\,& \frac{1}{|\beta|^\Delta}\,
\bigg(p^{\Delta(r+t+1-2k)}+2p^{\Delta(-r-t-1)}
\sum_{n=1}^\infty p^{-\Delta w n}
\bigg)\,\widehat{\phi}(x)\int_{Z_k} dy\,.
\end{align}
Moreover, from our choice \eqref{measureNormalization} of normalization for the integration measure, it follows that the various subsets have the following volumes:
\begin{align}
\int_{B} dy=\,&\frac{p^d-2}{p^d}\,|\beta|^{d}\,,
\nonumber \\[5pt] \nonumber
\int_{D_i} dy=\,&\frac{p^d-1}{p^d}\,|\beta|^{d}\,,
\\[5pt]  \label{volumes}
\int_{A_j} dy=\,&\frac{p^d-1}{p^d}\,p^{d(-r-1+j)}\,|\beta|^{d}\,,
\\[5pt] \nonumber
\int_{Z_k} dy=\,& \frac{p^d-1}{p^d}\,p^{d(-r-t-1+k)}\,|\beta|^{d} \hspace{5mm}\text{for }k> 0\,,
\\[5pt] \nonumber
\int_{Z_0} dy=\,& p^{d(-r-t-1)}\,|\beta|^{d}\,.
\end{align}
In the limit as $z$ approaches $x$ and $t$ tends to infinity, we see from \eqref{intPieces} and \eqref{volumes} that the integrals over $B$, $D_i$, and $A_j$ all tend to zero, while the integrals over $Z_k$ blow up as $p^{(\Delta-d) t}$. To obtain a finite limit, we must therefore first multiply the integral with a factor of $p^{(d-\Delta) t}$. But from the way we defined them, $t$ and $r$ depend on the function $\widehat\phi$ that provides the boundary conditions. However, the combination $r+t+1=s(z,C)$ is a purely geometric quantity that depends only on $z$, for it is the number of edges separating $z$ from the thermal circle. To investigate the limiting behaviour of \eqref{consideredIntegral}, we therefore first multiply $\mathcal{I}$ by $p^{(d-\Delta)s(z,C)}$, and for the sake of dimensional analysis we choose to also include a factor of $|\beta|^{\Delta-d}$, before we let $x$ tend to $z$:
\begin{align}
\lim_{x\rightarrow z}
|\beta|^{\Delta-d} p^{(d-\Delta)s(z,C)}\,
\mathcal{I}
=\,&
|\beta|^{\Delta-d}
\lim_{t\rightarrow \infty}
p^{(d-\Delta)(r+t+1)}
\sum_{k=0}^t 
\int_{Z_k} dy\,K^{(w)}_\Delta(z,y)\,\widehat{\phi}(y)
\nonumber \\[-8pt]
\label{limitAnswer}
\\[-8pt] \nonumber
=\,& \bigg(1+\frac{p^d-1}{p^d}\sum_{k=1}^{\infty}p^{-(2\Delta-d) k}\bigg) \widehat{\phi}(x)
=\frac{\zeta_p(2\Delta-d)}{\zeta_p(2\Delta)}\,\widehat{\phi}(x)\,.
\end{align}
From this equation, we see that \eqref{DirichletSolution} does indeed exhibit the limit \eqref{scaling} as advertised, which concludes our derivation of the solution to the Dirichlet problem.

We now turn to the boundary derivative. From the formulas we have provided, it can be straightforwardly verified that the difference between $|\beta|^{\Delta-d} p^{(d-\Delta)s(z,C)}\mathcal{I}$ and the right-hand side of \eqref{limitAnswer} tends to zero with a scaling $p^{-(2\Delta-d)t}$ as $t$ goes to infinity. To obtain a finite limit, we therefore multiply this difference by $p^{(2\Delta-d)s(z,C)}$. We wish, then, to determine the behaviour of the following difference as $z \rightarrow x$:
\begin{align}
\label{D}
\mathcal{D}=
|\beta|^{\Delta-d} p^{\Delta s(z,C)}\mathcal{I}
-
p^{(2\Delta-d)s(z,C)}\frac{\zeta_p(2\Delta-d)}{\zeta_p(2\Delta)}\,\widehat{\phi}(x)\,.
\end{align}
Now, from equation \eqref{intPieces} and \eqref{volumes}, one can show that when $\widehat{\phi}(y)=1$, the integral \eqref{consideredIntegral} evaluates to the following,
\begin{align}
\label{Iis1}
\int_{\partial T_{p^d}^{(w)}} dy\,K^{(w)}_\Delta(z,y)
=p^{(\Delta-d)s(z,C)}\frac{\zeta_p(2\Delta-d)}{\zeta_p(2\Delta)}
\bigg(1 - \frac{p^{-(2\Delta-d)s(z,C)}}{p^d}\frac{\zeta_p(\Delta)^2}{\zeta_p(\Delta-d)^2}\bigg)\,|\beta|^{d-\Delta}
\,.
\end{align}
Therefore, the second term in \eqref{D} can be re-expressed as\footnote{In the massless case, $\Delta = d$, and so it may appear that the subleading term in \eqref{reexpressed} is no longer suppressed in the $s(z,C)\rightarrow \infty$ limit. However, because of the factors of $\zeta_p(\Delta-d)$ in the denominator in \eqref{Iis1}, the subleading terms are actually identically zero in the massless case, and so the derivation we are here presenting of the normal derivative remains valid.}
\begin{align}
\label{reexpressed}
p^{(2\Delta-d)s(z,C)}\frac{\zeta_p(2\Delta-d)}{\zeta_p(2\Delta)}\,\widehat{\phi}(x)
=\frac{p^{\Delta s(z,C)}}{|\beta|^{d-\Delta}}
\int_{\partial T_{p^d}^{(w)}} dy\,K^{(w)}_\Delta(z,y)\,\widehat{\phi}(x)
+\mathcal{O}\big(p^{-(\Delta-d)s(z,C)}\big)\,.
\end{align}
The $z\rightarrow x$ limit of $\mathcal{D}$ can therefore be written as
\begin{align}
\label{almost}
\lim_{z\rightarrow x}\mathcal{D}=
\lim_{z\rightarrow x}
\frac{p^{\Delta s(z,C)}}{|\beta|^{d-\Delta}}
\int_{\partial T_{p^d}^{(w)}} dy\,K^{(w)}_\Delta(z,y)
\Big(
\widehat{\phi}(y)
-\widehat{\phi}(x)
\Big)\,.
\end{align}
For $y \in Z_k$ with $k\in \{0,...,t\}$, the integrand in \eqref{almost} is zero, so we can restrict the integration domain to the region $B\cup_i D_i\cup_j  B_j$. But for $y$ belonging to this region, it follows from the definitions of $K_\Delta(z,y)$ and $H_\Delta(x,y)$ given in Section~\ref{sec:Dirichlet} that 
\begin{align}
\label{HandK}
H_\Delta(x,y) = \frac{p^{\Delta s(z,C)}}{|\beta|^{\Delta}} K_\Delta(z,y)\,.
\end{align}
Applying \eqref{HandK} to \eqref{almost}, we therefore conclude that
\begin{align}
\lim_{z\rightarrow x}\mathcal{D}=
|\beta|^{2\Delta-d}
\int_{\partial T_{p^d}^{(w)}} dy\,H^{(w)}_\Delta(x,y)
\Big(
\widehat{\phi}(y)
-\widehat{\phi}(x)
\Big)\,,
\end{align}
which after a simple overall multiplication reproduces formula \eqref{boundaryDerivative} for the normal derivative.

\subsection{Two-point one-loop diagram}
\label{appendix:twoPoint}

In this section we provide intermediate steps to demonstrate how, by evaluating the bulk integral \eqref{D2int} for the two-point one-loop Witten diagram, one arrives at our answers in equations \eqref{D2oneloop} and \eqref{D2oneloopAlt}. We will consider in turn the case where the boundary points emanate from the same vertex on the thermal cycle and the case where they do not.

\subsubsection{Two-point one-loop diagram in the OPE regime}

\begin{figure}
    \centering
\begin{align*}
\begin{matrix}\text{
\includegraphics[scale=1]{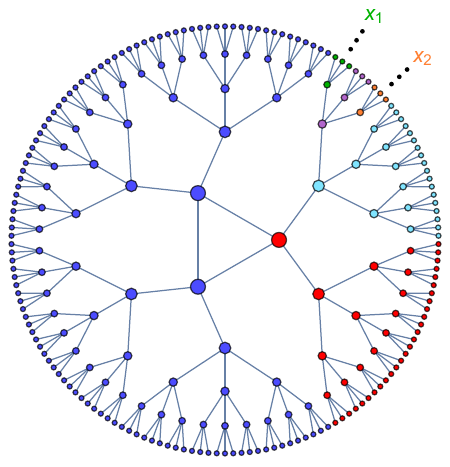}}
\end{matrix}
\end{align*}
    \caption{Depiction of the Tate curve using a color coding to partition the bulk space into separate contributions to the one-loop Witten diagram for the two-point correlator in the OPE regime. For this example, $p^d=w=3$ and $l=2$.}
    \label{fig:2ptDiagram}
\end{figure}

Assume $v(x_1)=v(x_2)$. We first break the bulk integral into six pieces $P_1$ to $P_6$ and pull out some overall factors common to these pieces:
\begin{align}
D_2^{(\text{1-loop, OPE})}
=
\frac{1}{|\beta|^{\Delta_{12}}}
\Big(P_1+P_2+P_3+P_4+P_5+P_6\Big)\,2\sum_{n=1}^\infty p^{-n\Delta_a w}\,.
\end{align}
The decomposition into six pieces is performed by partitioning the bulk space into six vertex sets as shown in Figure~\ref{fig:2ptDiagram}. To perform this decomposition we make use of the fact that we are currently considering a configuration of two boundary points $x_1$ and $x_2$ that emanate from the same vertex on the thermal cycle.

Explicitly summing over the contributions to the one-loop Witten diagram due to each of the six vertex sets, and recalling equation \eqref{lrelation}, which relates $|x_{1,2}|$ and $|\beta|$ to a number $l$, one finds that
\begin{align}
\textcolor{colour1}{P_1=} & \textcolor{colour1}{
\sum_{v=1}^{w-1}
\Big(p^{-v \Delta_1 }+p^{-\Delta_1 (w-v)}\Big)
\Big(p^{-v \Delta_2 }+p^{-\Delta_2 (w-v)}\Big)
\bigg(\sum_{n=0}^\infty p^{-n\Delta_1 w}\bigg)
\bigg(\sum_{n=0}^\infty p^{-n\Delta_2 w}\bigg)
}
\\
& \nonumber
\textcolor{colour1}{\bigg(1+(p^d-1)p^{-\Delta_{12aa}}
\sum_{n=0}^\infty p^{-n(\Delta_{12aa}-d)}
\bigg)\,,}
\\
\textcolor{colour2}{P_2=}\, & \textcolor{colour2}{
\bigg(1+2\sum_{n=1}^\infty p^{-n\Delta_1 w}\bigg)
\bigg(1+2\sum_{n=1}^\infty p^{-n\Delta_2 w}\bigg)
\bigg(1+(p^d-2)p^{-\Delta_{12aa}}
\sum_{n=0}^\infty p^{-n(\Delta_{12aa}-d)}
\bigg)\,,
}
\\
\textcolor{colour3}{P_3=} & \textcolor{colour3}{
\sum_{j=1}^{l-1}
p^{j\Delta_{12,aa}}
\bigg(1+2p^{-2j\Delta_1}
\sum_{n=1}^\infty p^{-n\Delta_1 w}
\bigg)
\bigg(1+2p^{-2j\Delta_2}
\sum_{n=1}^\infty p^{-n\Delta_2 w}
\bigg)
}
\\
&  \nonumber
\textcolor{colour3}{
\bigg(1+(p^d-1)p^{-\Delta_{12aa}}
\sum_{n=0}^\infty p^{-n(\Delta_{12aa}-d)}
\bigg)\,,
}
\\
\textcolor{colour4}{P_4=} & \textcolor{colour4}{
\,p^{l\Delta_{12,aa}}
\bigg(1+2p^{-2l\Delta_1}\sum_{n=1}^\infty p^{-n\Delta_1 w}\bigg)
\bigg(1+2p^{-2l\Delta_2}\sum_{n=1}^\infty p^{-n\Delta_2 w}\bigg)
}
\\
&  \nonumber
\textcolor{colour4}{
\bigg(1+(p^d-2)p^{-\Delta_{12aa}}
\sum_{n=0}^\infty p^{-n(\Delta_{12aa}-d)}
\bigg)\,,
}
\\
\textcolor{colour5}{P_5=} & \textcolor{colour5}{
\,p^{l\Delta_{12,aa}}
\sum_{j=1}^\infty
p^{-j\Delta_{1aa,2}}
\bigg(
1+2p^{-2(l+j)\Delta_2}
\sum_{n=1}^\infty
p^{-n\Delta_2 w}
\bigg)
\bigg(1+2p^{-2l\Delta_1}\sum_{n=1}^\infty p^{-n\Delta_1 w}\bigg)
}
\\
&  \nonumber
\textcolor{colour5}{
\bigg(1+(p^d-1)p^{-\Delta_{12aa}}
\sum_{n=0}^\infty p^{-n(\Delta_{12aa}-d)}
\bigg)\,,
}
\\
\textcolor{colour6}{P_6=} & \textcolor{colour6}{
\,p^{l\Delta_{12,aa}}
\sum_{j=1}^\infty
p^{-j\Delta_{2aa,1}}
\bigg(
1+2p^{-2(l+j)\Delta_1}
\sum_{n=1}^\infty
p^{-n\Delta_1 w}
\bigg)
\bigg(1+2p^{-2l\Delta_2}\sum_{n=1}^\infty p^{-n\Delta_2 w}\bigg)
}
\\
&  \nonumber
 \textcolor{colour6}{
\bigg(1+(p^d-1)p^{-\Delta_{12aa}}
\sum_{n=0}^\infty p^{-n(\Delta_{12aa}-d)}
\bigg)\,.
}
\end{align}
Evaluating the geometric series for each contribution and adding everything together, one recovers equations \eqref{D2oneloop}. Our claim in equation \eqref{KKzero} that the trivial geodesics do not contribute to $D^{\text{(1-loop)}}_2$ amounts in this regime to the statement that the sum over the six $P_i$ terms vanishes if we set $\Delta_a=0$.

\subsubsection{Two-point one-loop diagram outside the OPE regime}

\begin{figure}
    \centering
\begin{align*}
\begin{matrix}\text{
\includegraphics[scale=0.8]{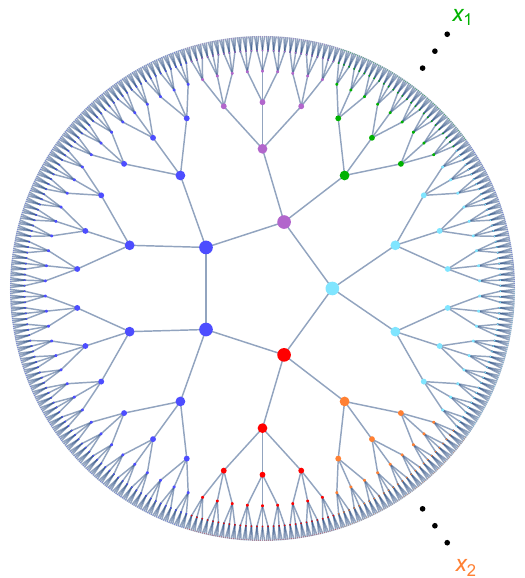}}
\end{matrix}
\end{align*}
    \caption{Depiction of the Tate curve using a color coding to partition the bulk space into separate contributions to the one-loop Witten diagram for the two-point correlator outside the OPE regime. For this example, $p^d=3$, $w=5$ and $u=2$.}
    \label{fig:2ptDiagramAlt}
\end{figure}

Assume now that $v(x_1)\neq v(x_2)$. Again, we split up the bulk integral into six pieces and pull out some shared overall factors:
\begin{align}
D_2^{(\text{1-loop, non-OPE})}
=
\frac{1}{|\beta|^{\Delta_{12}}}
\Big(Q_1+Q_2+Q_3+Q_4+Q_5+Q_6\Big)\,2\sum_{n=1}^\infty p^{-n\Delta_a w}\,,
\end{align}
where the integration domains corresponding to the six pieces are indicated in Figure~\ref{fig:2ptDiagramAlt}. These six pieces are given explicitly by
\begin{align}
\textcolor{colour6}{Q_1=}\, & \textcolor{colour6}{
\sum_{j=1}^{\infty}
\bigg(p^{j \Delta_1 }+2p^{-j\Delta_1} \sum_{n=1}^\infty p^{-wn\Delta_1}\bigg)
p^{-j\Delta_{2aa}}
\Big(
\sum_{n\in \mathbb{Z}}p^{-\Delta_2|u+nw|}\Big)
}
\\ \nonumber
&
\textcolor{colour6}{
\bigg(
1+(p^d-1)p^{-\Delta_{12aa}}\sum_{n=0}^\infty
p^{-n(\Delta_{12aa}-d)}
\bigg)\,,
}
\\
\textcolor{colour5}{Q_2=}\, & \textcolor{colour5}{Q_1\big|_{\Delta_1\leftrightarrow\Delta_2}\,,
}
\\
\textcolor{colour4}{Q_3=}\, & \textcolor{colour4}{
\Big(
\sum_{n\in \mathbb{Z}}p^{-\Delta_2|u+nw|}\Big)
\bigg(
1+2\sum_{n=1}^\infty p^{-w n \Delta_1}
\bigg)
\bigg(
1+(p^d-2)p^{-\Delta_{12aa}}\sum_{n=1}^\infty
p^{-n(\Delta_{12aa}-d)}
\bigg)\,,
}
\\
\textcolor{colour2}{Q_4=}\, & \textcolor{colour2}{Q_3\big|_{\Delta_1\leftrightarrow\Delta_2}\,,
}
\\
\textcolor{colour3}{Q_5=}\, & \textcolor{colour3}{
\sum_{v=1}^{u-1}
\sum_{n_1\in \mathbb{Z}}p^{-\Delta_1|v+n_1w|}
\sum_{n_2\in \mathbb{Z}}p^{-\Delta_2|u-v+n_2w|}
\bigg(
1+(p^d-1)p^{-\Delta_{12aa}}\sum_{n=1}^\infty
p^{-n(\Delta_{12aa}-d)}
\bigg)\,,
}
\\
\textcolor{colour1}{Q_6=}\, & \textcolor{colour1}{Q_5\big|_{u\rightarrow\, w-u}\,.
}
\end{align}
Adding together the six pieces results in the answer for the Witten diagram given in equation \eqref{D2oneloopAlt}. If we set $\Delta_a=0$, then, after formally evaluating all the geometric series, the sum of the six $Q_i$ terms vanishes in accordance with equation \eqref{KKzero}.

\subsection{Three-point tree diagram}
\label{appendix:threePoint}

In this section we demonstrate how the bulk three-point integral \eqref{OOO} can be evaluated and shown to equal the result given in equation \eqref{OOOanswer}.

\begin{figure}
    \centering
\begin{align*}
\begin{matrix}\text{
\includegraphics[scale=1.7]{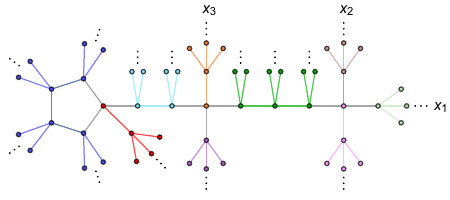}}
\end{matrix}
\end{align*}
    \caption{Partition of the Tate curve into various subsets for computation of the three-point correlator. For this figure, $p^d=3$, $w=5$, $\gamma=3$ and $\alpha=4$.}
    \label{fig:threePoint}
\end{figure}

By the isosceles property of ultrametric norms, of any three numbers $a$, $b$ and $a-b$, two of them must have the same norm, with the third having equal or smaller norm. This means that setting $a=x_1-x_3$ and $b=x_3-x_2$, we can assume without loss of generality that $|a|=|b|\geq |a+b|$ since this condition can always be attained by a relabelling of the $x_i$ boundary points. This choice amounts to assuming that the point of closest approach to the thermal cycle for the shortest geodesic between $x_1$ and $x_2$ is as far or farther from the thermal cycle than the point of closest approach for the shortest geodesic between $x_1$ and $x_3$, as in Figure \ref{fig:threePoint}.

From the way we defined the thermal norm $|\cdot|$ in Section~\ref{sec:Introduction}, the maximal value of $|x_1-x_3|$ is $|\beta|$. Therefore, under our assumption $|a|=|b|\geq |a+b|$, we may write
\begin{align}
|x_1-x_3|=|x_2-x_3|=|\beta|\,p^{-\gamma}\,,
\hspace{20mm}
|x_1-x_2|=|\beta|\,p^{-\alpha-\gamma}\,,
\end{align}
where $\alpha,\gamma \in \mathbb{N}_0$. Now, it turns out that the answer for the three-point correlator assumes a different functional form depending on whether $\alpha$ or $\gamma$ is equal to zero. We will focus on the most generic configuration, and so we take both $\alpha$ and $\gamma$ to be non-zero. For this configuration, the bulk space can be partitioned into nine vertex sets as show in Figure~\ref{fig:threePoint}. Accordingly, we can split up the bulk integral into nine pieces $\mathcal{P}_i$, which we can render dimensionless by pulling out an overall power of $|\beta|$,
\begin{align}
\sum_{z\in T_{p^d}^{(w)}}
K_{\Delta_1}(x_1,z)
K_{\Delta_2}(x_2,z)
K_{\Delta_3}(x_3,z)
=\frac{1}{|\beta|^{\Delta_{123}}}\sum_{i=1}^9 \mathcal{P}_i\,.
\end{align}
The nine pieces of the bulk integral are then given explicitly by
\begin{align}
\textcolor{colour9}{\mathcal{P}_1=} &
\textcolor{colour9}{
\bigg(1+(p^d-2)p^{-\Delta_{123}}\sum_{n=0}^\infty p^{-n(\Delta_{123}-d)}\bigg)p^{-\alpha\Delta_3}
\bigg(p^{\gamma\Delta_3}+2p^{-\gamma\Delta_3}\sum_{n=1}^\infty p^{-nw\Delta_3}\bigg)}
\\
& \nonumber
\textcolor{colour9}{
\bigg(p^{(\alpha+\gamma)\Delta_1}+2p^{-(\alpha+\gamma)\Delta_1}\sum_{n=1}^\infty p^{-nw\Delta_1}\bigg)
\bigg(p^{(\alpha+\gamma)\Delta_2}+2p^{-(\alpha+\gamma)\Delta_2}\sum_{n=1}^\infty p^{-nw\Delta_2}\bigg)\,,
}
\\
\textcolor{colour7}{\mathcal{P}_2=} &
\textcolor{colour7}{
\bigg(1+(p^d-1)p^{-\Delta_{123}}\sum_{n=0}^\infty p^{-n(\Delta_{123}-d)}\bigg)p^{-\alpha\Delta_3}
\bigg(p^{\gamma\Delta_3}+2p^{-\gamma\Delta_3}\sum_{n=1}^\infty p^{-nw\Delta_3}\bigg)
}
\\
& \nonumber
\textcolor{colour7}{
\bigg(p^{(\alpha+\gamma)\Delta_2}+2p^{-(\alpha+\gamma)\Delta_2}\sum_{n=1}^\infty p^{-nw\Delta_2}\bigg)
\sum_{j=1}^\infty p^{-j\Delta_{23}}
\bigg(
p^{(\alpha+\gamma+j)\Delta_1}
+2p^{-(\alpha+\gamma+j)\Delta_1}
\sum_{n=1}^\infty p^{-nw\Delta_1}
\bigg)\,,
}
\\
\textcolor{colour8}{\mathcal{P}_3=} &
\textcolor{colour8}{
\bigg(1+(p^d-1)p^{-\Delta_{123}}\sum_{n=0}^\infty p^{-n(\Delta_{123}-d)}\bigg)p^{-\alpha\Delta_3}
\bigg(p^{\gamma\Delta_3}+2p^{-\gamma\Delta_3}\sum_{n=1}^\infty p^{-nw\Delta_3}\bigg)
}
\\ & \nonumber
\textcolor{colour8}{
\bigg(p^{(\alpha+\gamma)\Delta_1}+2p^{-(\alpha+\gamma)\Delta_1}\sum_{n=1}^\infty p^{-nw\Delta_1 }\bigg)
\sum_{j=1}^\infty p^{-j\Delta_{13}}
\bigg(
p^{(\alpha+\gamma+j)\Delta_2}
+2p^{-(\alpha+\gamma+j)\Delta_2}
\sum_{n=1}^\infty p^{-nw\Delta_2 }
\bigg)\,,
}
\\
\textcolor{colour4}{\mathcal{P}_4=} &
\textcolor{colour4}{
\bigg(1+(p^d-2)p^{-\Delta_{123}}\sum_{n=0}^\infty p^{-n(\Delta_{123}-d)}\bigg)
\bigg(p^{\gamma\Delta_1}+2p^{-\gamma\Delta_1}\sum_{n=1}^\infty p^{-nw\Delta_1}\bigg)
}
\\
& \nonumber
\textcolor{colour4}{
\bigg(p^{\gamma\Delta_2}+2p^{-\gamma\Delta_2}\sum_{n=1}^\infty p^{-nw\Delta_2}\bigg)
\bigg(p^{\gamma\Delta_3}+2p^{-\gamma\Delta_3}\sum_{n=1}^\infty p^{-nw\Delta_3}\bigg)\,,
}
\\
\textcolor{colour5}{\mathcal{P}_5=} &
\textcolor{colour5}{
\bigg(1+(p^d-1)p^{-\Delta_{123}}\sum_{n=0}^\infty p^{-n(\Delta_{123}-d)}\bigg)
\bigg(p^{\gamma\Delta_1}+2p^{-\gamma\Delta_1}\sum_{n=1}^\infty p^{-nw\Delta_1}\bigg)}
\\
& \nonumber
\textcolor{colour5}{
\bigg(p^{\gamma\Delta_2}+2p^{-\gamma\Delta_2}\sum_{n=1}^\infty p^{-nw\Delta_2}\bigg)
\sum_{j=1}^\infty p^{-j\Delta_{12}}
\bigg(
p^{(j+\gamma)\Delta_3}
+2p^{-(j+\gamma)\Delta_3}
\sum_{n=1}^\infty p^{-nw\Delta_3}
\bigg)\,,
}
\\
\textcolor{colour6}{\mathcal{P}_6=} &
\textcolor{colour6}{
\bigg(1+(p^d-1)p^{-\Delta_{123}}\sum_{n=0}^\infty p^{-n(\Delta_{123}-d)}\bigg)
\bigg(p^{\gamma\Delta_3}+2p^{-\gamma\Delta_3}\sum_{n=1}^\infty p^{-nw\Delta_3}\bigg)
}
\\
& \nonumber
\textcolor{colour6}{
\sum_{j=1}^{\alpha-1}
p^{-j\Delta_3 }
\bigg(
p^{(\gamma+j)\Delta_1}
+2p^{-(\gamma+j)\Delta_1}
\sum_{n=1}^\infty p^{-nw\Delta_1 }
\bigg)
\bigg(
p^{(\gamma+j)\Delta_2}
+2p^{-(\gamma+j)\Delta_2}
\sum_{n=1}^\infty p^{-nw\Delta_2 }
\bigg)\,,
}
\\
\textcolor{colour3}{\mathcal{P}_7=} &
\textcolor{colour3}{
\bigg(1+(p^d-1)p^{-\Delta_{123}}\sum_{n=0}^\infty p^{-n(\Delta_{123}-d)}\bigg)
}
\\
& \nonumber
\textcolor{colour3}{
\sum_{j=1}^{\gamma-1}
\bigg(p^{j\Delta_1 }+2p^{-j\Delta_1 }\sum_{n=1}^\infty p^{-nw\Delta_1 }\bigg)
\bigg(p^{j\Delta_2 }+2p^{-j\Delta_2 }\sum_{n=1}^\infty p^{-nw\Delta_2 }\bigg)
\bigg(p^{j\Delta_3 }+2p^{-j\Delta_3 }\sum_{n=1}^\infty p^{-nw\Delta_3 }\bigg)\,,
}
\\
\textcolor{colour2}{\mathcal{P}_8=} &
\textcolor{colour2}{
\bigg(1+(p^d-2)p^{-\Delta_{123}}\sum_{n=0}^\infty p^{-n(\Delta_{123}-d)}\bigg)}
\\
& \nonumber
\textcolor{colour2}{
\bigg(1+2\sum_{n=1}^\infty p^{-nw \Delta_1 }\bigg)
\bigg(1+2\sum_{n=1}^\infty p^{-nw \Delta_2 }\bigg)
\bigg(1+2\sum_{n=1}^\infty p^{-nw \Delta_3 }\bigg)\,,
}
\\
\textcolor{colour1}{\mathcal{P}_9=} &
\textcolor{colour1}{
\bigg(1+(p^d-1)p^{-\Delta_{123}}\sum_{n=0}^\infty p^{-n(\Delta_{123}-d)}\bigg)\bigg(\sum_{n=0}^\infty p^{-nw\Delta_1 }\bigg)
\bigg(\sum_{n=0}^\infty p^{-nw\Delta_2 }\bigg)
\bigg(\sum_{n=0}^\infty p^{-nw\Delta_3 }\bigg)}
\nonumber \\
& 
\textcolor{colour1}{
\sum_{v=1}^{w-1}
\Big(p^{-v\Delta_1 }+p^{-(w-v)\Delta_1}\Big)
\Big(p^{-v\Delta_2 }+p^{-(w-v)\Delta_2}\Big)
\Big(p^{-v\Delta_3 }+p^{-(w-v)\Delta_3}\Big)\,.
}
\end{align}
Adding together the above nine pieces results in the following expression:
\begin{align}
\sum_{i=1}^9\mathcal{P}_i
=\,&f(\Delta_1|\Delta_2|\Delta_3)
\bigg(
p^{\alpha\Delta_{12,3}+\gamma\Delta_{123}}
\\ \nonumber
&
+2\zeta_p(w\Delta_1)p^{\gamma\Delta_{23,1}-w\Delta_1}
+2\zeta_p(w\Delta_2)p^{\gamma\Delta_{13,2}-w\Delta_2}
+2\zeta_p(w\Delta_3)p^{(\alpha+\gamma)\Delta_{12,3}-w\Delta_3}
\\ \nonumber
&
+2\zeta_p(w\Delta_2)\zeta_p(w\Delta_3)p^{-w\Delta_{23}}
+2\zeta_p(w\Delta_1)\zeta_p(w\Delta_3)p^{-w\Delta_{13}}
+2\zeta_p(w\Delta_1)\zeta_p(w\Delta_2)p^{-w\Delta_{12}}
\bigg)\,.
\end{align}
After reintroducing the dimensionful overall power of $|\beta|$ and re-expressing the result in a form that is invariant under simultaneous relabellings of the boundary points $x_i$ and scaling dimensions $\Delta_i$, one obtains the final answer \eqref{OOOanswer}.

{\setstretch{0.6}
\bibliographystyle{ssg}
\bibliography{literature}
}

\end{document}